# CIRCE: The Canarias InfraRed Camera Experiment for the Gran Telescopio Canarias

Stephen S. Eikenberry[†], Miguel Charcos[†], Michelle L. Edwards[†‡], Alan Garner[†], Nestor Lasso-Cabrera[†§], Richard D. Stelter[†], Antonio Marin-Franch[†§], S. Nicholas Raines[†], Kendall Ackley[†], John G. Bennett[†], Javier A. Cenarro[†§], Brian Chinn[†], H. Veronica Donoso[†], Raymond Frommeyer[†], Kevin Hanna[†], Michael D. Herlevich[†], Jeff Julian[†], Paola Miller[†], Scott Mullin[†], Charles H. Murphey[†], Chris Packham[†ᵀ], Frank Varosi[†], Claudia Vega[†], Craig Warner[†], A.N. Ramaprakash[ϕ], Mahesh Burse[ϕ], Sunjit Punnadi[ϕ], Pravin Chordia[ϕ], Andreas Gerarts[♣], Héctor de Paz Martín[♣], María Martín Calero[♣], Riccardo Scarpa[♣], Sergio Fernandez Acosta[♣], William Miguel Hernández Sánchez[♣], Benjamin Siegel[♣], Francisco Francisco Pérez[♣], Himar D. Viera Martín[♣], José A. Rodríguez Losada[♣], Agustín Nuñez[♣], Álvaro Tejero[♣], Carlos E. Martín González[♣], César Cabrera Rodríguez[♣], Jordi Molgó Sendra[♣], J. Esteban Rodriguez[♣], J. Israel Fernádez Cáceres[♣], Luis A. Rodríguez García[♣], Manuel Huertas Lopez[♣], Raul Dominguez[♣], Tim Gaggstatter[♣], Antonio Cabrera Lavers[♣], Stefan Geier[♣], Peter Pessev[♣], Ata Sarajedini[†], A.J. Castro-Tirado[♦]

[†]*Department of Astronomy, University of Florida, Gainesville, FL 32611, USA, eiken@ufl.edu*

[¥]*OTEG, National Oceanographic Center, European Way, Southampton, SO14 3ZH, UK*

[‡]*Large Binocular Telescope Obsercatory, 933 N. Cherry Ave, Room 552, Tucson, AZ 85721, USA.*

[§]*CEFCA, Centro de Estudios de Física del Cosmos de Aragón, Plaza San Juan 1 -2, Teruel, 44001 Teruel, Spain*

[ᵀ]*University of Texas - San Antonio, Department of Physics & Astronomy, University of Texas at San Antonio, One UTSA Circle, San Antonio, TX 78249, USA; National Astronomical Observatory of Japan, Mitaka, Tokyo 181-8588, Japan*

[ϕ]*IUCAA, Pune University Campus, Ganeshkhind, Pune, Maharashtra 411007, India*

[♣]*Gran Telescopio Canarias, Cuesta de San José, s/n - 38712 - Breña Baja - La Palma, España*

[♦]*Instituto de Astrofísica de Andalucía (IAA-CSIC), Glorieta de la Astronomía, s/n, 18008 Granada, España*

The Canarias InfraRed Camera Experiment (CIRCE) is a near-infrared (1-2.5 micron) imager, polarimeter and low-resolution spectrograph operating as a visitor instrument for the Gran Telescopio Canarias 10.4-meter telescope. It was designed and built largely by graduate students and postdocs, with help from the UF astronomy engineering group, and is funded by the University of Florida and the U.S. National Science Foundation. CIRCE is intended to help fill the gap in near-infrared capabilities prior to the arrival of EMIR to the GTC, and will also provide the following scientific capabilities to compliment EMIR after its arrival: high-resolution imaging, narrowband imaging, high-time-resolution photometry, imaging polarimetry, low resolution spectroscopy. In this paper, we review the design, fabrication, integration, lab testing, and on-sky performance results for CIRCE. These include a novel approach to the opto-mechanical design, fabrication, and alignment.
*Keywords*: near-infrared; imaging; diamond-turned optics; polarimetry; Gran Telescopio Canarias

## 1. CIRCE Overview

The University of Florida (UF) has developed the Canarias InfraRed Camera Experiment (CIRCE) as a near-infrared open-use visitor instrument for the Gran Telescopio Canarias (GTC) 10.4-meter telescope. CIRCE is an all-reflective 1-2.5 μm imager with a 3.4x3.4-arcminute field-of-view, offering imaging polarimetry and fast photometry as well as built-in upgrade paths for narrow-band imaging and low-resolution grism spectroscopy. While CIRCE was built in the University of Florida Infrared Instrumentation Laboratory, CIRCE is the first UF

"training instrument", built primarily by graduate students, a postdoctoral fellow, and the Principal Investigator, and with the advice and support of the UF instrumentation team. This approach provided a path for graduate students and young PhDs to be trained as instrument builders, which is not generally possible in the context of modern major facility instrumentation projects for large telescopes, where the much larger instruments, budgets, and teams generally prevent untrained personnel from "learning the ropes" of instrument-building in a broad sense.

CIRCE development took place primarily at UF, as well as some partner institutions. The University of Florida provided approximately $1.2M in funds and resources to allow the completion and commissioning of CIRCE. This was augmented by an NSF grant to support graduate student participation, and the Spanish International School for Advanced Instrumentation provided support for additional students/postdocs for some activities. The Inter-University Center for Astronomy and Astrophysics (IUCAA) provided the electronic/software for interface to the detector array controller. Finally, while the GTC did not provide any direct funds or resources for CIRCE development, the close collaboration of the GTC staff and the provision of the Folded Cassegrain (FC) focal station for CIRCE were absolutely crucial to the completion of the instrument.

In this paper, we review the design, fabrication, integration, lab testing, and on-sky performance results for CIRCE on the GTC. We begin in Section 2 with the science drivers and requirements for CIRCE. In Section 3, we present the CIRCE opto-mechanical system, followed by the cryo-vacuum system in Section 4. In Sections 5 and 6 we present the electronics and software aspects of CIRCE. We present the key laboratory test results for CIRCE in Section 7. Finally, in Section 8 we present results from the on-sky performance verification for CIRCE on the GTC in 2015.

## 2. CIRCE Science Drivers & Requirements

The CIRCE Science Advisory Committee (S.S. Eikenberry, A.J. Castro-Tirado, E. Mediavilla, A. Sarajedini, M. Tapia) developed the key science cases for the CIRCE instrument prior to its design, and from this we extracted the key science requirements for CIRCE. We present those preliminary science cases and their resulting drivers in this section.

### 2.1. *Relativistic jet formation in microquasars*

The discovery of black-hole/relativistic-jet X-ray binaries in our own Galaxy -- called the ``microquasars" for their analogous relationship to the black-hole-powered jets in quasars -- is revolutionizing studies of relativistic jets. The archetypal microquasars, GRS 1915+105 and GRO J1655-40, exhibit collimated jets with apparent superluminal motions of $v_{app}$ ~1.25c, and true velocities of ~0.9c (Mirabel & Rodriguez, 1994; Tingay et al., 1995; Hjellming & Rupen, 1995). Furthermore, both of these systems appear to be stellar X-ray binaries powered by accretion onto a black hole -- GRO 1655-40 is a binary system with a F-star secondary and a 7 $M_\odot$ compact primary (Orosz & Bailyn, 1997), and GRS 1915+105 contains a K/M III secondary and a 14 $M_\odot$ compact object primary (Greiner et al., 2001). Not only are these black hole/relativistic-jet systems much closer to us than their AGN cousins, allowing more detailed study of their properties and structure, but they also vary much more rapidly -- with timescales from milliseconds to hours, rather than years to decades. Thus, the Galactic microquasars offer tremendous potential as ``test laboratories" for probing the black-hole/relativistic-jet connection.

Simultaneous IR and X-ray observations of GRS 1915+105 have already shed considerable new light on the question of how compact objects form relativistic jet outflows. In many epochs, the X-ray lightcurves displayed large repeating X-ray dips, revealing the disappearance and apparent re-filling of the inner accretion disk, while high-speed continuous-frame IR images showed corresponding large-amplitude IR flares (see Eikenberry et al., 1998a; Mirabel et al., 1998; Fender & Pooley, 1998; Eikenberry et al., 2000; Rothstein et al., 2005; Eikenberry et al., 2008, and references therein), consistent with synchrotron emission from ejected synchrotron-emitting plasma at relativistic speeds (Mirabel and Rodriguez, 1994), this then leads to a picture where the inner portion of an accretion disk is being swept up and ejected from the system in the form of relativistic jets (Eikenberry et al., 1998a).

The high-speed photometric mode of CIRCE will be crucial for carrying out such further observations. In addition, the queue-scheduled mode of operating GTC will enable much better IR coverage (of flaring behaviors in GRS 1915+105, for example) when coupled with Target of Opportunity ``triggers'' from monitoring observations at X-ray and radio wavelengths. Another important capability provided by CIRCE will be linear polarization measurements of the IR synchrotron flares. Many theoretical models predict significant polarization of the jets, and measurement of this polarization will constrain the models and provide important insight into the make-up of the ejecta (e.g., are they primarily electron-proton or electron-positron material?). While radio observations show no significant polarization of most flares, Faraday de-rotation across the jet itself could wipe out any intrinsic polarization signature at these wavelengths -- an effect which will be relatively insignificant at IR wavelengths.

## 2.2. *Emission-line surveys for massive star clusters*

Massive stars are extremely rare -- for a Salpeter Initial Mass Function (IMF), solar-mass stars (0.5-2 $M_\odot$) outnumber high-mass stars (>20 $M_\odot$) by approximately 100,000 to 1. However, these stars exert a disproportionate effect on their host galaxies, dominating: mechanical energy input into the interstellar medium (via winds and supernova explosions); radiated luminosity (either direct emission or far-infrared from heated dust); ionizing UV radiation; chemical enrichment of CNO elements (via supernovae). These stars are also crucial for understanding the origin of compact objects (such as microquasars), which are born in the core-collapse death events of massive stars. Despite their importance, relatively little is known about the population of these stars in our Galaxy, in part due to their locations in the crowded Galactic Plane and relatively high extinction (both local and along the line-of-sight). As a result, optical study has apparently missed the vast majority of massive star clusters in the Milky Way, and new space-based mid-infrared surveys such as GLIMPSE and WISE are demonstrating that this is true. Identification and study of these clusters can thus led to much greater insight into the star formation history and evolution of our own Galaxy. Furthermore, the massive stars in these clusters produce the compact objects powering both microquasars (above) and γ-ray bursts (below), and their study in these environments is already leading to important insights into the formation and evolution of compact objects (i.e. SGR 1806-20 (Eikenberry et al., 2001) and the X-ray sources in the Antennae (Clark et al., 2005)).

We plan to use CIRCE to carry out emission-line imaging surveys for such massive star clusters in our Galaxy using narrow-band filters centered on lines which are diagnostic of massive stellar atmospheres (Brγ (2.156 μm), Paβ (1.28 μm), HeI (2.058 μm), HeII (2.112 μm), and FeII (1.64 μm)). Due to the high extinctions expected, these lines will be much more powerful probes than any optical transitions. Given the expected sensitivity of CIRCE, for 1% line filters we can survey a given field in 3 lines (plus 2 narrow-band continuum filters) to a depth allowing 5σ detection of a dwarf OB star with typical line strengths (~5 Angstroms) at a distance of 10 kpc with extinction $A_K$ ~3 mag in ~1-hour of wall-clock time. We will then use low-resolution follow-up spectroscopy with CIRCE to confirm the status of these stars and to provide more detailed classification of the massive stellar type. In addition, higher-resolution follow-up spectroscopy (for instance, with ``the upcoming MIRADAS spectrograph on GTC with R>20,000 - Eikenberry et al., 2014) can be used to determine the metallicity of the stars and their environments (providing insight into the chemical enrichment status and history of the Milky Way, and by extrapolation other galaxies) as well as the intrinsic stellar luminosity from terminal wind velocities (which in turn leads to distance measurements and thus to the 3-dimensional star-forming structure of the Milky Way).

### 2.3. *Transient observations: GRBs & gravitational wave events*

Gamma-ray bursts (GRBs) have remained a puzzle for many astrophysicists since their discovery in 1967. Only recently has it been possible to carry out deep multi-wavelength observations of the counterparts associated with the long-duration GRBs class just within a few hours of occurrence, thanks to the observation of the fading X-ray emission that follows the more energetic γ-ray photons once the GRB event has ended. The fact that this emission (the afterglow) extends to longer wavelengths led to the discovery of optical/IR/radio counterparts starting in 1997, greatly improving our understanding of these sources.

CIRCE/GTC can significantly contribute to this field by means of target of opportunity (ToO) imaging/polarimetry of gamma-ray bursts (GRBs) and X-ray flashes (XRFs) afterglows localized by satellites like SWIFT. GTC is operated in a queue-mode the majority of the time, allowing the straightforward implementation of such ToO observations on timescales of a few hours. Fixed-time late epoch observations will be also performed in order to study the GRB host galaxies and determine, together with optical observations, the spectral energy distribution. Such observations are even more important in the era of the advanced gravitational wave detectors.

### 2.4. *Deep JHK photometry of Galactic globular clusters*

Very few Galactic globular clusters (GGC) possess sufficiently deep JHK color-magnitude diagrams to allow a meaningful comparison with the theoretical models of metal-poor stellar populations (as compared to about one-half with optical photometry). As such, we know very little about where these models break down in the low-metallicity, low-mass regime. In addition, since many calibrations between $(V-K)_0$ and effective temperature (used to derive chemical abundances from spectra) rely on the theoretical models to some extent, assessing the validity of these models is important. To remedy this situation, we will use CIRCE to image GGCs with declinations north of -30 degrees and within ~30 kpc down to ~4 mag below the main sequence turnoff. This will provide an unprecedented look at the lower main sequence of these objects in V-(JHK). The combination of sensitivity and large FOV of CIRCE on GTC make this work feasible. These projects will require exposures of ~4 hours per field for photometric depth required to reach the un-evolved main sequences of GGC (S/N=10 at Ks = 22.4 mag).

### 2.5. *Distance scale of Local Group dwarf galaxies*

Grocholski & Sarajedini (2002) present a calibration that yields the absolute K-magnitude of the red clump ( MK(RC) ). For a significant range of age and metallicity, the value of MK(RC) is a constant. As such, we can combine it with apparent K(RC) to get distance. The red clump is easily identified in various stellar populations including the low surface brightness dwarf galaxies in the Local Group. Many of these galaxies also contain other standard candles such as RR Lyrae variables and Cepheids. The red clump distances will be used as a check and a calibrator of these other distance methods. In addition, since many of the dwarf galaxies contain dust, the near-IR has the potential to provide more robust distances as compared with optical observations. The photometric depth required (S/N ~20 at Ks < 20 mag) is not prohibitive, requiring only ~10 minutes of integration per field with CIRCE/GTC.

### 2.6. *Metal abundance gradients in M31 and M33*

Tiede et al. (2004) have an empirical calibration between red giant branch slope in K, J-K and metallicity for populations older than about 1 Gyr. We are interested in imaging M31 and M33 fields along the major and minor axes, constructing K vs. J-K color-magnitude diagrams, and using our calibration to study the run of metal abundance with position in the disk, halo, and bulge of these two spiral galaxies. We would then like to compare these numbers with the Milky Way. The metallicity gradient itself and the differences between spiral galaxies with a range of masses can tell us about the early epochs of star formation in spirals. The near-IR is advantageous here because it is less susceptible to extinction which is a significant problem in spirals. To facilitate this study, we require ~20 radial fields in M31 and ~10 fields in M33 with a S/N ~20 at Ks ~22 mag which is approximately 1 magnitude above the helium burning red clump. Thus, we will need roughly 40 hours of on-source time for M31 and 20 hours for M33. This amounts to about ~10 nights of total observing time with CIRCE on GTC (allowing for standard star exposures and other calibration frames).

**2.7.** *Key CIRCE science requirements*

The following table summarizes the key science requirements for CIRCE, as derived from the science cases above.

**Table 1 - CIRCE Key Science Performance Requirements**

| Parameter | Value | Comment |
|---|---|---|
| Wavelength range | 1-2.5 µm | Maximum sensitivity near-IR windows from the ground |
| Detector | HAWAII-2RG | Detector of choice for this bandpass |
| Pixel scale | 0.1-arcsec/pixel | Good sampling of best seeing at GTC (~0.3-arcsec FWHM); Required for crowded field science |
| Field of View | 3.4x3.4-arcmin | Driven by pixel scale and detector format |
| Optical system | All-reflective | High throughput for best sensitivity |
| Polarimetry | Linear | Required for relativistic jet science case; Combined with Canaricam [ref], provides 1-13 µm polarimetry on GTC |
| Fast Photometry | >10 Hz | Required for relativistic jet science case |
| Spectroscopy | R~500 | Required for massive stars science case |

## 3. CIRCE Opto-Mechanical System

**3.1.** *Primary optics*

The CIRCE primary opto-mechanical design uses a novel approach to all-reflective metal mirror systems employing off-axis diamond-turned aspheric mirror figures. We initially chose this approach for the combination of high throughput and excellent image quality it offers. However, there are also some challenges associated to this approach, we describe below. In this section, we describe the CIRCE optical design, its fabrication, testing, and alignment, and on-telescope performance.

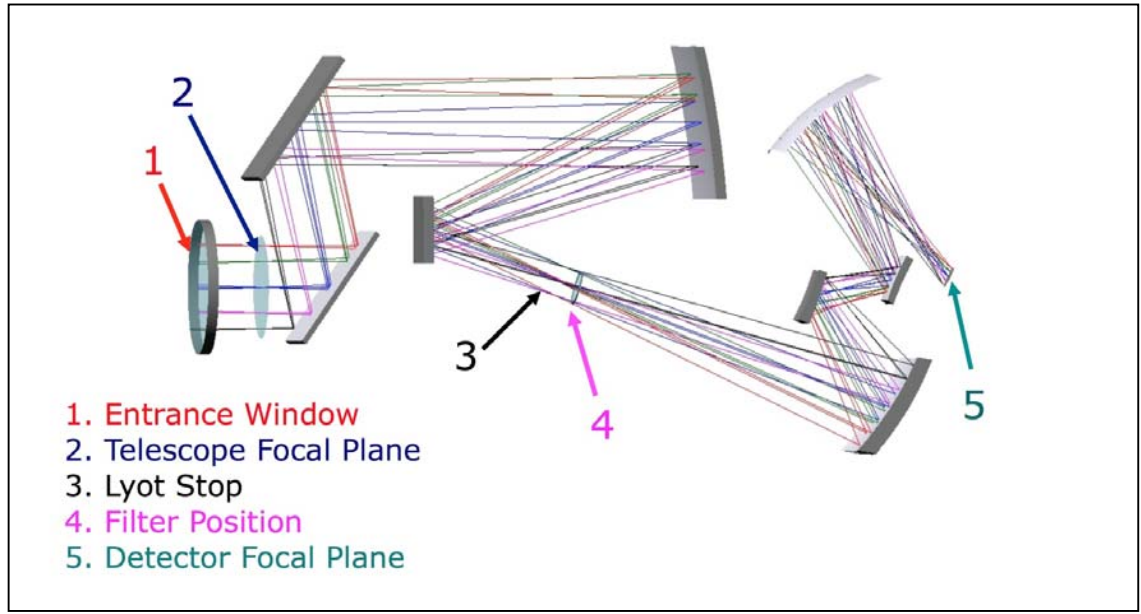

Figure 1 - Schematic layout of the CIRCE optical system

The basic optical design for CIRCE was carried out by J.M. Rodgers of Optical Research Associates, in consultation with the CIRCE team. We considered several basic design options, all of which consisted of a collimator/camera optical relay which re-images the telescope focal plane on the detector, with an intermediate pupil image of the telescope secondary mirror as a convenient location for filters, cold stops, grisms, Wollaston prisms, etc. The final design consisted of two fold mirrors to maintain the optical system inside of the GTC Folded Cassegrain instrument envelope, two aspheric mirrors (one prolate ellipsoid and one hyperboloid) comprising the collimator, and a 4-mirror system comprising the camera. Three of the camera mirrors are aspheres with conic sections (one prolate ellipsoid, one hyperboloid, and one oblate ellipsoid), while the final mirror has a 6th-order polynomial aspheric surface. We present the layout of this system in Figure 1, the optical prescription in Table 2, and the some typical nominal spot diagrams in Figure 2.

**Table 2 - CIRCE Optical Prescription**

| Surface | Radius of Curvature (mm) | Conic Constant | X-width (mm) | Y-width (mm) | Following thickness (mm) | Comments |
|---|---|---|---|---|---|---|
| Focal Plane | -1822.515 | 0 | 170 | 170 | 130.000 | GTC focal plane |
| Fold 1 | Inf | 0 | 184 | 228 | 279.000 | |
| Fold 2 | Inf | 0 | 204 | 252 | 790.982 | |
| Coll Mirror 1 | -1466.21 | -0.3354 | 252 | 250 | 534.230 | |
| Coll Mirror 2 | -2771.27 | -47.55 | 108 | 110 | 256.531 | |
| Pupil image | Inf | 0 | 53.4 | 53.4 | 679.000 | Cold stop location |
| Cam Mirror 1 | -453.72 | -0.5811 | 180 | 168 | 208.860 | |
| Cam Mirror 2 | -322.22 | -65.107 | 74 | 90 | 128.302 | |
| Cam Mirror 3 | 423.53 | 13.710 | 70 | 82 | 305.338 | |
| Cam Mirror 4 | 387.80 | 0.0527 | 182 | 204 | 397.912 | 4th order coeff = $-8.178 \times 10^{-11}$; 6th order coeff = $-2.576 \times 10^{-16}$ |
| Detector | Inf | 0 | | | | HAWAII-2RG |

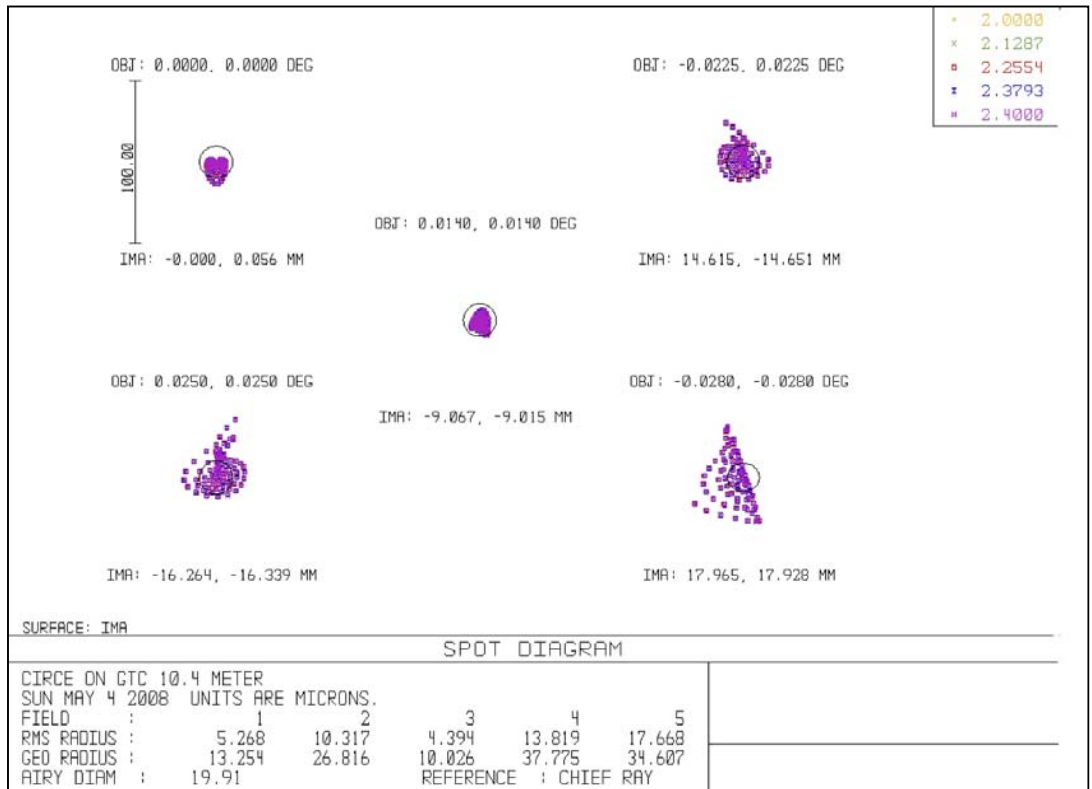

Figure 2 - Typical spot diagrams representing nominal CIRCE optical performance over the field of view. We note that for many field positions, the majority of the rays fall within one Airy Disk and are tightly packed. The corners show the most aberration, but no one particular type dominates and the spot sizes are still small, if irregularly shaped.

Using modern precision diamond-turning techniques, such surfaces are relatively straightforward to generate (to rough precision) in 6061-Al mirror blanks such as those used for CIRCE, with generally very good "straight-off-the-machine" surface roughness. However, testing mirrors with such a large departure from the best-fitting sphere (in order to make the needed final corrections to the mirror shape) is extremely difficult and quite expensive. Initial quotations from vendors indicated that the metrology costs alone would exceed the diamond-turning cost of the mirrors by a factor of more than three.

Furthermore, even with perfectly-fabricated mirrors, such a highly-aspheric design is also very sensitive to alignment errors. The CIRCE team carried out optical tolerancing analyses in collaboration with ORA (for details, see Edwards et al., 2008) to determine the resulting requirements on mirror alignment. We found that CIRCE could provide excellent image quality over its full field of view, but only if we kept mirror decenters to <37 microns RMS in all dimensions. While such alignment tolerances are certainly possible to achieve, this is again significantly complicated using standard alignment techniques on a mirror-by-mirror basis due to the highly aspherical surface shapes of the individual mirrors. Furthermore, this tolerance included all errors introduced by placing the mirrors - including the mirror pin locations to their brackets and the bracket pin locations to the optical bench, and the bench pin locations relative to the overall coordinates.

In response to this, the CIRCE team developed a novel approach to the CIRCE mirror fabrication and alignment, in collaboration with E. Stover (then at Janos Technology, Keene, NH). This solution was based on four important facts:

> (i) The construction of the CIRCE mirrors and bench from identical 6061-AL material ensures that a successful alignment of the optical system at room temperature will hold at operating temperatures (~80K) due to homologous contraction of the entire system.
>
> (ii) While single-pass generation of the optical surfaces cannot guarantee meeting the performance requirements for a system such as CIRCE in the individual mirrors, it can and does regularly produce optical surfaces which are relatively close to the final shape.
>
> (iii) While standard fabrication/alignment techniques cannot guarantee meeting the alignment accuracy requirements for CIRCE without adjustment (requiring metrology for feedback), standard fabrication of mirrors/brackets can and does produce highly repeatable placement/alignment (both on the optical bench and in the diamond-turning machine).
>
> (iv) While optical metrology of highly-aspheric individual mirrors is very difficult and expensive, measuring the optical characteristics of mirror *systems* - especially collimator and camera systems - made up of such mirrors is actually straightforward and inexpensive.

The resulting solution is as follows. First, we fabricate each of the mirrors using best practices in diamond-turning, with precision dowel pins for highly-repeatable placement onto their mounting brackets. In parallel, we fabricate the mirror brackets and CIRCE optical bench. We then place an entire optical subsystem (either the collimator or the camera) on the bench (with its support system), with the mirrors in their final brackets, and test the optical performance of this subsystem. This test involves a standard interferometric measurement of the subsystem using a double-pass approach with a flat mirror at the focal plane to retro-reflect the light. This produces a map of aberrations in the system, but with no knowledge of exactly *where* the aberrations/errors exist (i.e. which mirror is out of specification).

However, as long as the mirrors are all relatively close to their nominal surface figure, we do not necessarily need this information. Instead, we can use the optical metrology/design software to project the entire aberration map onto a single surface of our selection in the system. This then determines a set of corrections which, when applied to that mirror, will correct the overall performance of the system. This is essentially a static version of the technique used to correct atmospheric wavefront errors in astronomical adaptive optics.

Once the corrections are determined, the "correction mirror" can be removed from its bracket and placed (with good repeatability) into the diamond-turning machine, where corrective cuts are applied. The corrected mirror can then be replaced onto the optical bench, and the subsystem-level (camera/collimator) test above can be repeated to verify final performance.

The mechanical support of the bench and the primary optics constituted another key design decision for the CIRCE system. This support system must hold the primary optical system of CIRCE within the alignment tolerances above while experiencing the full range of gravity vectors at the GTC Folded Cassegrain focal station, maintain good overall alignment of the CIRCE optics with the GTC telescope optics, and also be compatible with good thermal conduction from the $LN_2$ cryogen reservoir.

We considered several options for the support system, before settling on a dual "tank beam" approach for CIRCE. In this case, a pair of parallel, long (roughly scuba-tank-shaped) $LN_2$ tanks bolt directly to the optical bench underside. The tanks each act as hollow cylindrical mechanical beams, providing vastly greater stiffness to deflection than the bench alone. At the same time, they hold the $LN_2$ and provide good thermal contact between the cryogen and the primary CIRCE optics. Both tank beams attach to a half-circle shaped aluminum bulkhead which in turn bolts to the G10 thermal isolation system described in Section 4 below.

The fabrication and alignment of the primary CIRCE optics followed the approach above. Vulcan Machine (Tampa, FL) fabricated the optical bench, with the CIRCE team providing cryogenic strain relief of the bench at several stages during the fabrication process. The University of Florida Physics and Astronomy Machine Shop fabricated the mirror brackets and mirror blanks. The same shop also fabricated the tank beams, taking special care to preserve good dimensionality of the beam mechanical interfaces, to avoid introducing stress-induced deformations into the optical bench,. We then delivered the entire primary optics assembly (tank beams, bench, mirrors, and brackets) to Lightworks Optics (who were subcontracted by Janos for this work when Erik Stover changed employers to Lightworks). After completing the fabrication process, Lightworks Optics delivered the aligned primary optics assembly to the CIRCE team at UF for testing (see Figure 3).

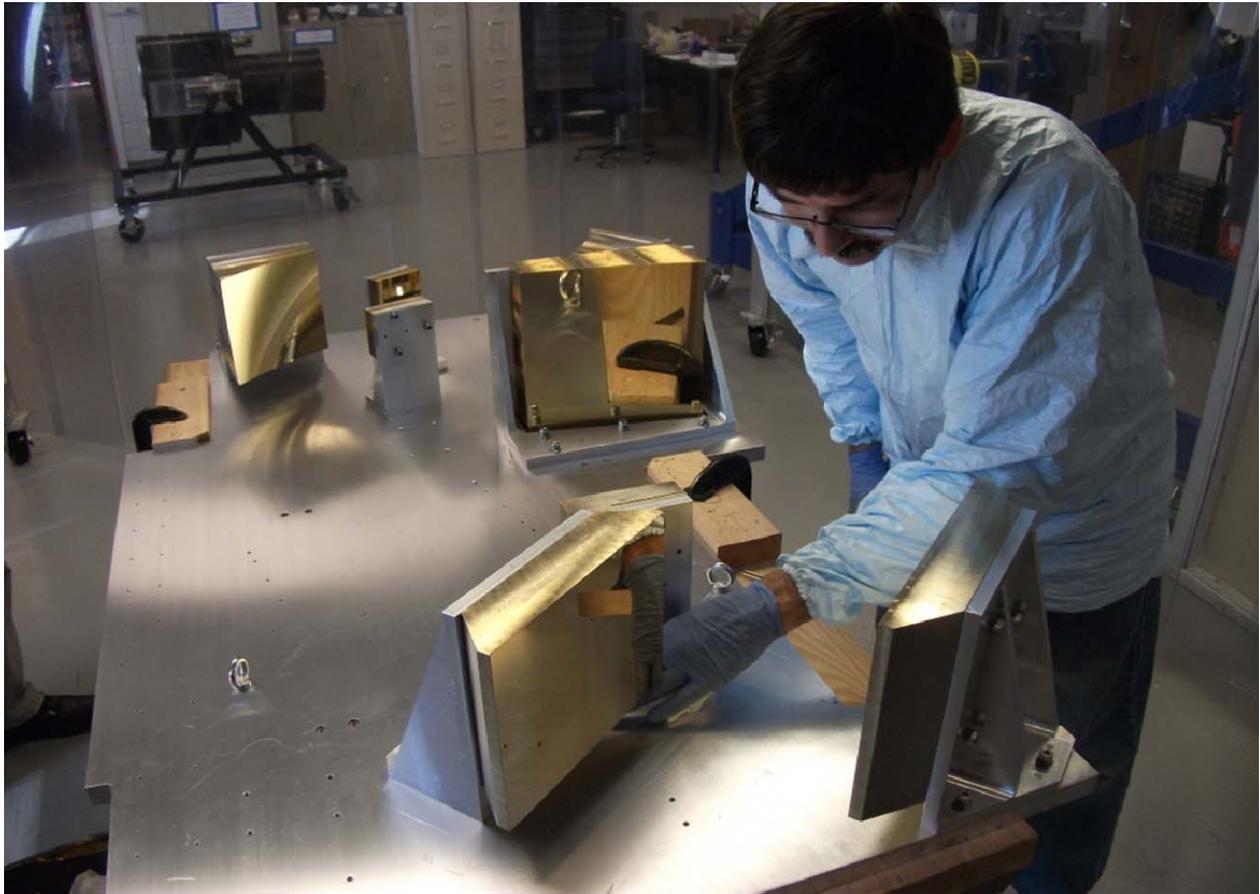

Figure 3 - CIRCE team member Nick Raines inspecting the bench/mirror assembly after delivery from the manufacturer. The fold mirrors are in the photo foreground, with the camera mirrors in the background.

### 3.2. CIRCE polarimetry subsystems

Polarimetry in CIRCE is performed using a selectable rotating half waveplate (HWP) and a selectable $MgF_2$ Wollaston prism beam splitter. We follow the solution proposed by Oliva (1997) using a Wedged Double-Wollaston (WeDoWo) that measures the linear Stokes parameters with a single exposure. This capability is important for rapidly-varying sources (such as microquasars) where the intrinsic changes in flux/polarization are not necessarily slow compared to the timescale for changing the HWP orientation, for instance. In practice, the observation consists of two exposures with the HWP at angles 0º and 45º. In this configuration, we expect to obtain an accuracy of 0.15% in the measurement of Stokes parameters for SNR of 700:1 (Clarke et al 1983; Simmons & Stewart 1985).

Images from the telescope are located in the focal plane between the entrance window and the first fold mirror. The HWP is inserted close to the focal plane where the beam size is minimal. The focal plane mechanism (described in detail below) contains masks for polarimetry and carries the HWP rotation mechanism. The HWP covers the polarimetry aperture. The beam splitter is a cemented Wedged Double-Wollaston (WeDoWo), which is located in the grism wheel near the pupil plane in the collimated beam. This configuration enables simultaneous measurements of the three linear Stokes parameters with a single observation. The WeDoWo (Oliva 1997) consists of two $MgF_2$ 27.5x55-mm$^2$ prisms dividing the pupil intensity into four beams with angles about $+1.5\delta_{eo}/+0.5\delta_{eo}/-0.5\delta_{eo}/-1.5\delta_{eo}$, with polarization main axes at 0°/90°/+45°/-45° and width ~25mm. Bernhard Halle Nachfolger GmbH provided both of these components for CIRCE, with the Wollaston prism being a custom design, which we describe below.

The angular separation of two beams by a Wollaston prism depends on its material and geometry as $\delta_{eo}=2\Delta n \tan\Omega$, with $\Omega$ the prism angle and $\Delta n=(n_e-n_o)$ the material birefringence. Then, the field of view (FOV) is given by the separation on the sky of both rays following $206265*D_p D_{tel}^{-1}\delta_{eo}$ (arcsec), where the pupil and telescope diameter ($D_p$ and $D_{tel}$) are in the same units. However, the prism effect of the WP creates elongation of the image proportional to the separation of the rays. The lateral chromatism for a particular wavelength in the range between $\lambda_1$ and $\lambda_3$ is quantified in terms of the parameter $V = \Delta n(\lambda)/(\Delta n(\lambda_3) - \Delta n(\lambda_1))$ that depends on the birefringence of the material at $\lambda$, $\lambda_1$ and $\lambda_3$. The third major factor to consider is the reflection at the prism interface that depends on the index of refraction. Low refraction index material accepts higher input angles. This factor and the opto-mechanical properties are major factors in the selection of the material. Once the material is chosen the optimization of the polarimetric system is determined by the geometry of the WeDoWo and the size of the polarimetric aperture. Two angles of the WeDoWo affect the separation of the beam: the angle of the edge ($\beta$) splits rays between both sectors of the Wollaston prism; the angle of the prism ($\alpha$) defines the separation of the e-o rays in each Wollaston prism comprising the WeDoWo. The WeDoWo can be coupled with FSi to decrease the chromatism. This layer defines a third angle ($\theta$) of the geometry. Once the material of the WeDoWo is defined the geometry of the system for a given FOV is calculated to first order from Snell's laws for each ray. However, CIRCE is a complex system with aspheric mirrors and accurate analytical results are complex. We explain in Charcos et al. (2008) how we obtain optimum results in the FOV from simulations and the chromatic aberrations from ZEMAX computation.

Hough et al (1994) discuss materials in the NIR range for Wollaston prisms. In selecting a material, considerations must be given to its transmission, required beam separation, birefringence of the material, refractive index and wavelength dependence. The most common and cheapest material in this wavelength range is $MgF_2$. It provides good optical transmission below 6μm wavelengths, with typical absorption of ~0.04 cm$^{-1}$ at 2.7μm. $MgF_2$ has indices of refraction $n_o$=1.3836, $n_e$=1.3957 at 0.4μm with reflection losses at the 2 surfaces of about 5.2% at 0.6μm. In addition, it is moderately achromatic for NIR range with $n_o$ ($n_e$) ranging from 1.37964 (1.38521) at 1μm to 1.36000 (1.37060) at 3μm. In addition, its thermo-optical properties adapt well to cryogenic conditions - at 0.4μm, the variation of the index of refraction is $dn_o/dT=2.3\times10^{-6}$ and $dn_e/dT=1.7\times10^{-6}$ corresponding to a variation of the deviation angle of 0.11%. On the other hand, thermal expansions are $a_a$=13.7x10$^{-6}$/K; $a_c$=8.48x10$^{-6}$/K, corresponding to a variation of the expansion of 0.07mm for the CIRCE pupil size. The crystal is not significantly deformed, but special optical cement is required for deployment at liquid nitrogen temperatures.

The HWP is a 45x45mm$^2$ octagon made of MgF$_2$ and quartz crystal. It has a retardation of λ/2±4% in the range 700-2500nm. Because its thermal expansion coefficient is 5.5x10$^{-7}$ K$^{-1}$ the contraction of the HWP is on the order of 6µm. Both plates are cemented with Infrasil 302 to avoiding breaking at low temperatures. This configuration results in low chromatic aberrations even though the thickness of the HWP introduces defocus in the beam.

### 3.3. *CIRCE spectroscopy design*

In addition to polarimetry, the combination of a focal plane mask and a pupil-space mechanism allows for an easy upgrade of CIRCE to provide grism spectroscopy. The CIRCE focal plane mask for polarimetry will also include slits for spectroscopy. These will include slits with widths of 0.3, 0.4, 0.6, and 0.8-arcsec width. All slits will have lengths of 10-arcsec to allow on-slit AB nodding for sky background subtraction.

We have modeled CIRCE's spectroscopic performance using the FLAMINGOS grisms (Elston et al., 2003). For this ZEMAX analysis, we placed the FLAMINGOS grisms at the grism wheel location of CIRCE which is also in the collimated beam near the pupil image. The results showed an expected resolution of 950 for a 0.3-arcsec width slit, and 450 for a 0.6-arcsec width slit – well-matched to the vision of CIRCE being a low-resolution spectrograph. We deployed the grisms in CIRCE during 2016.

### 3.4. *Entrance window*

CIRCE uses a 280-mm diameter Infrasil entrance window with anti-reflection coating on both sides. We considered using CaF$_2$ for this, due to its improved thermal conductivity compared to Infrasil (to prevent condensation from radiative cooling of the window). However, due to both the cost and birefringence of CaF$_2$, we chose Infrasil.

### 3.5. *Focal plane mechanism*

The CIRCE focal plane mechanism carries masks for polarimetry and spectroscopy, as well as the rotating half-wave plate (HWP) optic. The short back focal distance of the GTC FC focal station combined with the need for a cryogenic vacuum environment for CIRCE placed significant constraints on the design of this mechanism. We considered several design solutions before arriving at the final design we present here. The base design is a large aluminum frame with a linear translation stage of 180x250-mm. The stage rides on two stainless steel rails with Frelon linear sliding bearings. It holds a plastic lead screw nut mated to a 1/4-20 dual-lead screw directly coupled to the shaft of a Phytron VSS-42 motor. The linear translation stage moves between two limit switches intended to define the full range of travel. In addition, in case of switch failure resulting in driving the stage into its hard limits, the plastic nut is designed to provide the first failure point (thus protecting the more expensive components such as the Phytron motor from damage). When retracted from the beam, the stage leaves the full imaging field of view unvignetted.

The HWP rotation mechanism also attaches to the translation stage and "rides" on that stage into and out of the optical beam. The HWP optics and its holder are permanently fixed in front of the polarimetric mask locations of the translating stage. In this way, when being used the HWP is the first optic inside of the CIRCE entrance window, providing maximum modulation (and thus calibration) of the instrumental polarization signature. The original design of the HWP mechanism used an experimental steel belt drive to achieve rotation of the HWP. However, due to reliability issues, that mechanism was replaced in early 2016 with a spur gear mechanism.

### 3.6. *Pupil mechanism*

The CIRCE pupil mechanism consists of 5 wheels of 9.375-in diameter. The wheels were CNC-machined from a single piece of cryogenically stress-relieved 6061-T6 aluminum alloy, with a central hole for the mechanism hub, five filter locations, and an outer ring. The wheels ride on a central hub shaft and have gear teeth cut on the outer diameters to allow helical spur gear drive by cryogenic stepper motors. The wheels attach to the hub shaft via commercial radial bearings (440C Stainless steel, ABEC-7), and the hub shaft itself is made from ground 303 stainless steel. The wheels are cooled by being pre-loaded against caged pin thrust bearings that are placed between each wheel. The housing is made of 6061-T6 Al alloy, and has its surfaces acid-etched and anodized.

The stepper motors are Portescap motors prepared for cryogenic use, with 303 stainless steel drive gears and shafts. The wheel drive gear placement is designed for proper engagement at cryogenic temperatures, with shaft/hub pre-loading and compliance that also allows operation at ambient temperature for pre-cooldown testing. A fixed hub piece provides an anchor point for the disc washers that provide the axial preloading.

The first four wheels have five positions - four allocated for holding optical components, and one open position to pass light. Three of the wheels are filter wheels, providing a total of 12 filter positions. One wheel, located at the location of the pupil image, contains masks including a circular pupil stop. The final "grism" wheel has four positions - one open and three larger positions designed for holding spectroscopic grisms and the polarimetric Wollaston prism. Note that we took one of the 5-slot wheels and turned it into the Wolly/grism wheel – the 4-slot wheel was refitted for filters (in fact we needed special filter holders made to deal with it).

An import design decision here was the use of a fixed circular mask for the Pupil stop for CIRCE. GTC has a non-circular pupil image, matching the 6-fold symmetry of the mirror segments. Due to the alt-az nature of the GTC telescope, the pupil image will rotate about the optical axis as the instrument tracks a target field across the sky. Thus, in order to optimally match the pupil image, the Pupil stop would need a rotation mechanism to keep a quasi-hexagonal coldstop rotationally aligned with the pupil image. While such mechanisms can be constructed, may be needed for mid-infrared instruments (see e.g. Telesco et al., 2008), it represents a substantial complication and cost driver for near-infrared instruments such as CIRCE. Thus, we investigated the effectiveness of using a simple, fixed, circular coldstop with a diameter selected to optimize the signal-to-noise ratio for CIRCE, as well as the requirements for the alignment of such a circular coldstop with the pupil image. We found the following key points:

> (i) For background-limited observations, the optimal diameter for a fixed circular coldstop is 0.995 times the GTC pupil image diameter (which is the minimal circular diameter which produces no vignetting).
>
> (ii) With an optimally-sized circular fixed cold stop, the CIRCE sensitivity will only degrade by < 1% overall at a nominal ambient temperature (T=275K), and will only degrade by < 2% at warm temperatures (T=285K). All of these degradations are referenced to a perfectly aligned rotating coldstop which exactly matches the GTC pupil image shape.
>
> (iii) Pupil decenters of < 0.06 of the pupil image diameter produce < 2% sensitivity degradation, and pupil decenters of < 3.7% of the diameter produce < 1% sensitivity degradation.

### 3.7. *Detector mount*

The CIRCE detector mount is the standard HAWAII-2RG detector mount provided by Teledyne Imaging Systems. We designed a simple kinematic mounted aluminum bracket (using the same material as the bench, to eliminate differential thermal contraction) to attach the detector mount to the CIRCE optical bench at the desired detector location. We then attach a tilted conical baffle to the detector mount, in order to block stray light scattered from other mirror surfaces in the optical system. The SIDECAR array control electronics are mounted on a similar bracket attached to the cold bench, with a short flex cable connection between the two.

### 4. CIRCE Cryo-Vacuum System

The CIRCE cryo-vacuum system provides a cryogenic vacuum operating environment for the primary optical system, detector, and mechanisms, as well as the mechanical interface between the primary opto-mechanical system and the Gran Telescopio Canarias. Here, we review the primary components of this system.

#### 4.1. *Vacuum jacket*

The CIRCE jacket comprises a 1.08-meter diameter cylindrical volume with a 1.75-meter length as prescribed by the maximum envelope constraint at the GTC Folded Cassegrain focal station. The cylinder body consists of a telescope interface flange, which attaches to the GTC FC flange and also connects internally to the CIRCE thermal isolation system. The remainder of the vacuum jacket cylindrical volume consists of two longer cylinders. Our original design called for a single cylinder, but we switched to two shorter cylinders for relative ease of handling during assembly and servicing of the instrument. The vacuum jacket includes a front lid, with the entrance window opening, and a rear lid with vacuum penetrations for the vacuum pumpout, vacuum gauge, cable feedthrough panels, and $LN_2$ fill ports (all described below).

#### 4.2. *Thermal isolation subsystem*

The CIRCE thermal isolation system connects the cold mass of the instrument rigidly to the CIRCE vacuum jacket telescope interface flange, while at the same time providing good thermal isolation between them. Previous subsystems we have built typically use a thin-walled G10 cylinder fabricated by "rolling" thin strips of G10 into the appropriate diameter and thickness. This cylinder of G10 is then typically epoxied into grooves cut into Al6061-T6 endcaps which provide bolted attachment points to the telescope interface flange and the cold mass, respectively. However, in order to avoid significant flexure of the cold mass with respect to the telescope interface (and thus the telescope optical axis), we calculated that the G10 thickness required for CIRCE would be so large that we would not be able to roll it into the desired diameter. We investigated buying manufactured G10 tubes as well as having one custom-fabricated externally, but both options were difficult and expensive.

Thus, we developed an alternate design consisting of flat G10 sheets arranged in a hexagonal pattern as shown in Figure ABC. This structure allows us to use monolithic G10 sheets, which are commercially available in suitable thickness/dimensions, are cryo-rated, and are very inexpensive. We then create welded metal endcaps, with the "cold side" being aluminum and the "warm side" stainless steel. By placing steel mounting pads on the steel side, we further increase the thermal path length through the support structure and improve the thermal isolation. The individual G10 pieces are cut to size and epoxied into place as before (albeit without any composite layering of strips, and without bending or stressing the G10 sheets).

#### 4.3. *Liquid nitrogen subsystem*

As described above, the CIRCE $LN_2$ "tank beams" form an integral part of the opto-mechanical support system of the instrument. They attach to the cold bulkhead at their ends, as well as to the optical bench via machined flats on their top surfaces. We designed the tank beam wall thickness to provide sufficient thermal conduction path to avoid major thermal gradients in the instrument - this is especially crucial in the Folded Cassegrain environment where CIRCE operates, as the changing gravity vector will cause the liquid cryogen to move substantially inside of each tank. We chose a two-tank system for CIRCE in part to avoid major "sloshing" of liquid cryogen inside the instrument in the short dimension.

Each tank beam has a fill port which penetrates the radiation shields and attaches via stainless steel bellows to an O-ring interface to the vacuum jacket. The fill port end is threaded to allow the installation of a hollow stainless steel "stinger" tube in each tank beam. The end point of the stinger is located at the geometric center of the tank beam. In this way, as long as each tank is less than 1/2-full, the stinger end is always above the liquid level. This allows vapor from cryogen boiloff to escape, while at the same time preventing liquid cryogen leakage - regardless of the instrument orientation. Thus, CIRCE is purposely designed to be "full" at the halfway point of each tank (a fill volume of about 27 liters). This has a theoretical design hold time of approximately 40 hours, and our experience in the laboratory and at GTC is that the hold time is typically ~30-34 hours (depending on ambient temperature and its associated radiative thermal load). This allows straightforward daily $LN_2$ fills by GTC staff - a pre-requisite for CIRCE operations on the GTC.

### 4.4. *Radiation shields*

The CIRCE radiation shields consist of two aluminum sheet metal shells welded to solid aluminum interface flanges for attachment to the structure. The inner shell (the "active shield") is thermally coupled to the CIRCE cold mass, providing a dark environment for the detector/optics. The outer shell (the "passive shield") is mechanically attached to the CIRCE cold structure, but thermally isolated via G10 standoffs. Thus, the passive shield is only radiatively coupled to the vacuum jacket and the cold mass. Both the shields are wrapped in multiple layers of aluminized mylar "super insulation" to reduce emissivity, resulting in a significantly minimized radiative thermal load on the cold mass of CIRCE.

### 4.5. *Vacuum feedthroughs*

In CIRCE, the vacuum jacket provides good thermal and vacuum isolation, however it hinders the introduction of electrical signals inside the dewar. Signals to/from temperature sensors, warmup heater, motors, and science array then need to pass through the hermetically sealed vacuum shield, and the thermally isolated cold shield before arriving to the different components. The solution used in CIRCE to cross both shields consists of two concentric apertures of 304.8 mm and 254 mm in the vacuum and cold shield, respectively. These apertures allow access to the dewar interior and are hermetically closed with o-ring sealed plates at the vacuum interface. Four hermetic pass-through D-sub connectors attached to the cold shield plate and four hermetic pass-through MIL-C-M83723 Series-III connectors attached to the vacuum plate allow the housekeeping signals to enter into the dewar without losing thermal or vacuum isolation. The hermetic D-sub connectors mounted on the connectors cold plate facilitate the connection of the cables by using the standard mate D-sub connectors. However, the Series-III connectors mounted on the connector vacuum plate can only be connected to the mate connectors in the external side of the plate. The internal side of the Series-III connectors presents long pins to connect the cables, i.e., non-standard connectors. To solve the lack of standard connectors in this side of the plate, a fanout board is attached to all the pins of the Series-III connectors. The fanout board redirects all the signals from the pins to standard D-sub connectors mounted on the same fanout board. This solution allows the use of standard connectors at the same time that increase the stiffness of the entire system.

## 5. CIRCE Electronics Systems

All of CIRCE's electronics are commanded from the GTC control room observer computer through the Local Control Unit (LCU) computer. The CIRCE LCU and other key components of all subsystems are housed in a temperature controlled enclosure located beside the dewar at the Folded Cassegrain focal station. Cabling is used to connect the enclosure to the instrument. CIRCE installed on the telescope at GTC is shown in Figure 4.

CIRCE's electronics consist of three major subsystems - mechanism control, housekeeping, and detector control. The mechanism control electronics allow the observer to control the different movable parts of the instrument.

The function of the housekeeping electronics is to maintain the instrument working under cryo-vacuum conditions. The science array is independently controlled by the detector electronics.

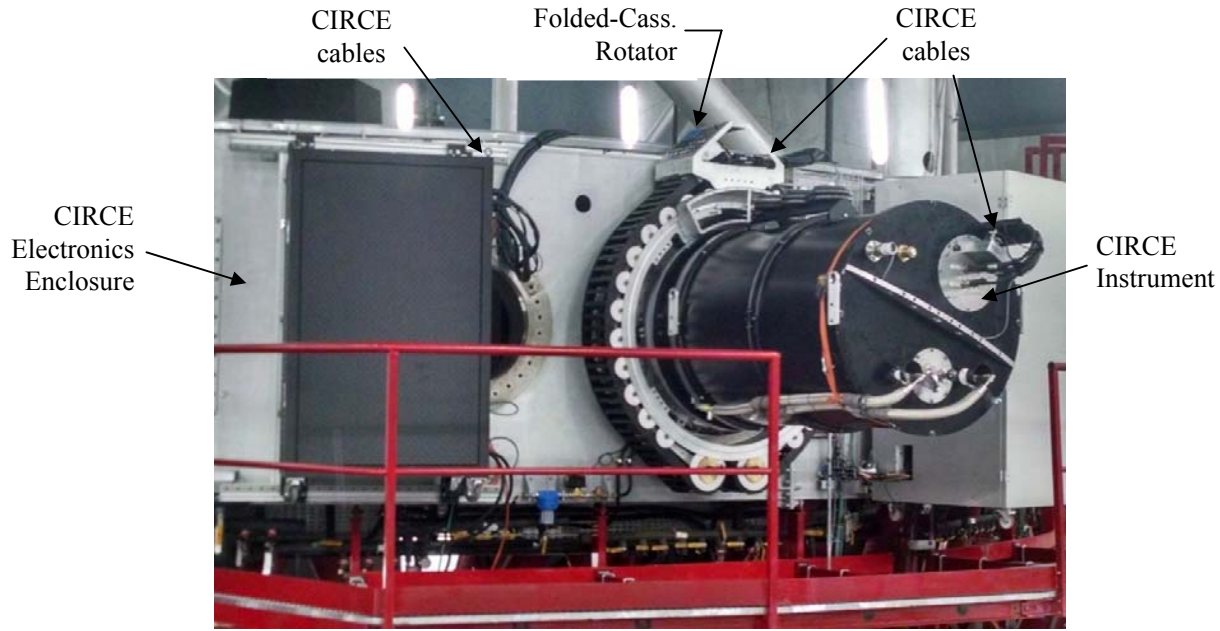

Figure 4 - CIRCE installed on the GTC Folded Cassegrain Rotator.

In the following sections, we describe in detail each individual component of the electronic design. Section 5.1 describes the mechanism control electronics, and in Section 5.2 we describe the housekeeping electronics. In Section 5.3, we cover the detector control electronics. Finally, we provide the description of the CIRCE electronics thermal enclosure, LCU, and the communications between the control computer (observer computer) and the different systems of the instrument in section 5.4.

**5.1.** *Mechanism control electronics*

The mechanism control electronics are responsible for the movement of the different parts of the instrument. All motors used in CIRCE (except the focal plane mechanism) are Portescap P532 stepper motors which have been degreased to avoid freezing of the grease and tested at cryogenic temperature to ensure proper operation. The motors work in conjunction with limit and/or home switches that indicate the end of the allowed movement in a direction or a reference position for the circular movements, respectively. All motors and switches are connected to Intelligent Motion Systems IM483I Microstepping indexer/drivers. These are located inside the motor drive chassis that sits inside the electronics thermal enclosure, and can be disconnected individually or by blocks to facilitate the testing of the different systems.

The motor drive chassis contains all the electronic components necessary for running the stepper motors used in CIRCE. Eight (8) Intelligent Motion Systems IM483I indexers, one per motor, sit inside the motor drive chassis driving the signals to and from the motors and switches through the motor cables. The indexer/drivers feed the A+, A-, B+, and B- power phases of the motors, and receive the signals of the limit and/or home switches. Communication with the indexers is though RS232 serial ports. All of the indexers/drivers are powered by two (2) 40V power supplies that are protected with a 10 Amp circuit breaker. A fan is used to draw the heat generated by the indexer/drivers and power supplies out of the motor chassis.

In the CIRCE instrument there are two types of mechanisms: Rotary and Linear. Rotary mechanisms use a single limit switch as a homing position switch and the linear mechanisms use two limit switches, one at each end of travel. As described above, the focal plane mechanism is a linear translation stage that allows the introduction of the different spectroscopic slits and/or the polarimetric half wave plate (HWP) in front of the light beam. The focal plane mechanism incorporates one motor to linearly translate the stage into and out of the light path. Two limits switches indicate when the stage is totally in and totally out. The HWP is a rotary mechanism utilizing one motor, with one home switch to indicate a reference position for the rotation. Meanwhile, the filter box is made up of five (5) rotary mechanisms that includes: 3 filters wheels, one Wollaston/grism wheel, and one Lyot wheel. Each wheel is moved by a motor, and referenced by a home switch. Finally, the third component of the moving mechanism system is the window cover mechanism. It's a linear mechanism that moves a protection cover plate in front of the entrance window when the instrument is not operational. The cover plate is moved by a single motor with two limit switches – in and out.

### 5.2. *Housekeeping electronics*

The housekeeping electronics are essential to the proper operation of the CIRCE instrument. The housekeeping electronics includes all the electronic components necessary for monitoring the instrument while operating at the working temperatures and pressures of the system. Besides being responsible for monitoring the pump down, cooldown, and warmup processes of the instrument, the housekeeping electronics are also responsible for monitoring and controlling the temperature of the most delicate part of the instrument, the Hawaii-2RG science array. The temperature and pressure electronics of CIRCE provide continuous monitoring over the temperature and pressure inside the dewar.

CIRCE provides continuous monitoring of the internal dewar pressure through a Pfeiffer Vacuum gauge PKR-251, externally attached to the dewar. The vacuum gauge is directly connected to a Pfeiffer Pressure Monitor TPG-261 located inside the temperature and pressure chassis that sits inside the electronics thermal enclosure. The housekeeping electronics continuously monitor the temperature of the optical bench through eight (8) 2N2222 diode temperature sensors. These temperature sensors are distributed around the optical bench and directly connected to the Lakeshore Temperature Monitor LS-218, located inside the temperature and pressure chassis that sits inside the electronics thermal enclosure.

The HAWAII-2RG science array has the lowest heat transfer rate of all the components of the instrument, which means that it is the last component to reach the working temperature during the cooldown process as well as the last component to warm up. For that reason, the housekeeping electronics has a section exclusively dedicated to monitoring and controlling the temperature of the science array. The temperature of the array is automatically controlled through a Lakeshore Cryogenic Temperature Controller LS-331 located inside the temperature and pressure chassis that sits inside the LCU rack (Sec. 7.3). Connected to the Lakeshore LS-331, two temperature sensors along a small custom heater are dedicated to the control of the temperature in the science array.

As mentioned above, the temperature and pressure inside the dewar are monitored and controlled through the temperature & pressure chassis inside the thermal enclosure. Three (3) different components sit inside this chassis: Lakeshore Cryogenic Temperature Controller LS-331, Lakeshore Temperature Monitor LS-218, and a Pfeiffer Pressure Monitor TPG-261. The temperature controller LS-331 has two channels for temperature sensor inputs, and one channel for warmup heater power output, and is dedicated to temperature stabilization of the science array. The temperature monitor LS-218 is an eight (8) temperature sensor inputs used to monitor the temperature in the dewar interior. The pressure monitor TPG-261 communicates with the pressure gauge connected to the dewar to monitor the pressure inside the dewar. Each one of those components has an independent circuit breaker for protection, and an on/off state LED lights.

CIRCE's housekeeping electronics also include a warmup heater system. Details of the design process are described in Lasso-Cabrera (2012). The warmup system can accelerate the warmup process by introducing heat directly into the dewar using 5 in-parallel Ohmite warmup heaters. Each heater consists of 3 in-series 155 Ohms cartridges, providing 104 Watts per heater for a total of 520 Watts. The warmup process is controlled through the warmup box, which is externally attached to the dewar and provides the power to feed the warmup heaters. An Omega CN8551-RTD-DC1 Temperature Controller and an Ohmitrol Solid-State Power Control Switch allow for manual control of the total power transferred to the dewar through the warmup heater cable. The warmup process is protected with a SELCO module located inside the dewar. The SELCO module automatically disconnects the warmup heaters when the temperature inside the dewar rises above 300K, avoiding overheating of the instrument.

### 5.3. *Detector control electronics*

CIRCE incorporates readout electronics based on the commercial SIDECAR ASIC system provided by Teledyne Imaging Systems. The SIDECAR ASIC is a Focal Plane Array (FPA) controller, fully compatible with the H2RG and other image arrays (Loose et al., 2003b, 2006, 2007), that allows in-Dewar operation with low power consumption and high noise immunity. The commercial SIDECAR ASIC system provided by Teledyne connects on one side directly with the H2RG and on the other side with a JADE2 card – replaced by the IUCAA system in CIRCE – that acts as a USB-2 interface between the SIDECAR ASIC and the acquisition computer. The JADE2 card is designed to work at room temperature and is connected to the cryogenic SIDECAR ASIC through a single 15 inch flex cable also provided by Teledyne. This means that the connection between the SIDECAR ASIC, which is directly connected to the science array at cryogenic temperature, and the JADE2 card, which is located outside the dewar at room temperature, has to be done with a single 15 inch flex cable – no intermediate connectors – to guarantee proper operation. The use of a single cable to connect the SIDECAR ASIC and JADE2 card hinders its integration on cryogenic system, such as CIRCE, where all the electronic signals entering or leaving the dewar pass through the cold and vacuum shield by means of hermetic connectors. The solution to this (developed by Teledyne and GL Scientific for another project originally) is to build a custom connector consisting of an aluminum plate that seals against the dewar via an o-ring, with a groove in the center for passing the 15 inch flex cable through. Once the cable is introduced in the groove, cable and groove are potted with a special hermetic epoxy, creating a hermetically sealed custom connector. This unique solution provides a custom connector to pass through the vacuum shield and is replicated to pass the flex cable through the cold shield. UF fabricated the parts for this connector, and GL Scientific assembled and potted the cable. CIRCE uses two 15-inch cables in series (one potted and one not) to connect the SIDECAR ASIC to its dewar-external control interface. The Hirose connections are mechanically fixed using custom-designed 3D-printed PET plastic clamps.

Besides the flex cable problem, the JADE2 card presents two other limitations for its integration into the design of CIRCE: first, the JADE2 card limits some of the functionalities of the SIDECAR ASIC controller, e.g., the possibility of capturing high rate subframes, thus preventing the creation of a fast photometry mode; and second, the JADE2 card only works under the Windows operating system, which is not a standard in astronomical observatories. The limitations presented by the SIDECAR ASIC - JADE2 system for its incorporation into CIRCE led the CIRCE team to find an alternative solution for the readout electronics. IUCAA has developed a fully functional data acquisition system called IUCAA SIDECAR Drive Electronics Card (ISDEC) which replaces the Teledyne JADE2 interface card (Ramaprakash et al., 2010). The ISDEC board is also being used for the Robert Stobie Spectrograph (RSS-NIR) being built at the University of Wisconsin for the Southern African Large Telescope (SALT). The ISDEC board connects to the SIDECAR ASIC, allowing the entire system to work under the Linux operating system, also allows the capture of high rate subframes, thus overcoming some of the limitations found when using the JADE2 card.

**5.4. *CIRCE electronics enclosure***

As stated previously, the CIRCE electronics enclosure is located beside the Dewar and that it contains all the electronics systems necessary for connecting, monitoring, and controlling the different components of the instrument from the GTC observer computer. All the functions of the CIRCE Instrument are controlled by the CIRCE Local Control Unit (LCU) in the electronics enclosure.  We used the computer provided by IUCAA with the I-SDEC system described above as the LCU.  The LCU contains the software that runs the motors, controls the temperature of the science array, and monitors the temperature and pressure of the dewar interior.  Two Ethernet networks cards are installed into the LCU. One is connected to the GTC public network. The GTC observer computer remotely commands GANESH through the observatory primary network. For safety purposes, an observatory secondary network is directly connected to a MOC 100BT Network Switch which is a requirement of the GTC and that was provided by the GTC.  The second network card in the LCU is directly connected to an Ethernet Switch that allows the creation of an internal private network which isolates the different components of the LCU from external commands.  Also connected to the LCU is the ISDEC.  LCU communication with the ISDEC is through two (2) USB 2.0 extenders cables daisy chained together.  Two (2) cables were required to span the distance between the LCU and the ISDEC. The electrical enclosure is shown in Figure 5.

 Connected to the private network Ethernet switch is a BayTech RPC3A-16 AC Power Control Module and a PERLE CS9000 16 Serial Port Terminal Server which allows commands of the components listed below

- (8) IM483I Microstepping indexer/drivers
- Lakeshore 331
- Lakeshore 218
- TPG-261

The remote management port of the PERLE CS9000 and the Baytech RPC3A-16 are connected to the GTC-provided MOC for management and diagnostics in case of failure.

CIRCE electronics enclosure is connected directly to the observatory UPS power. A power disconnect tagout/lockout switch is used to provide safety while working in the electronics enclosure.  The LCU and the BayTech RPC3A-16 Power Control Module are powered directly from the observatory power.  All other electronics are connected through the BayTech which allows remote control over the power of the components directly connected to it.

The GTC requires that any electrical enclosure/device that resides in the telescope dome cannot emit more than 150W of heat into the dome.  To meet this requirement the electrical enclosure is lined with 1" rigid insulation and the internal temperature of the enclosure is controlled with two liquid to air heat exchangers and a temperature sensor.  One Thermatron 721 heat exchanger and one Thermatron 723 heat exchanger cool the cabinet. The system pumps cold air to the top of the enclosure through ducts located at the sides of the racks, and pulls the air down and through the electronics to cool them. The heat exchangers are part of Cooling Circuit IV at GTC - combined they receive 20.4 l/min of coolant at $5^O$C.  They are run in parallel, with a pressure drop of 0.96 bar from input to output, which meets the GTC requirements.  The coolant supply is split between the two exchangers, and the 723 unit removes approximately 2000 W while the 721 unit removes approximately 1000 W.  Therefore the net power removed from the cabinet is approximately 3000 W, which far exceeds the estimated maximum of 1000 W to be dissipated by the electronics. An electro-valve is included in accordance to GTC guidelines to prevent unnecessarily heating of the coolant.

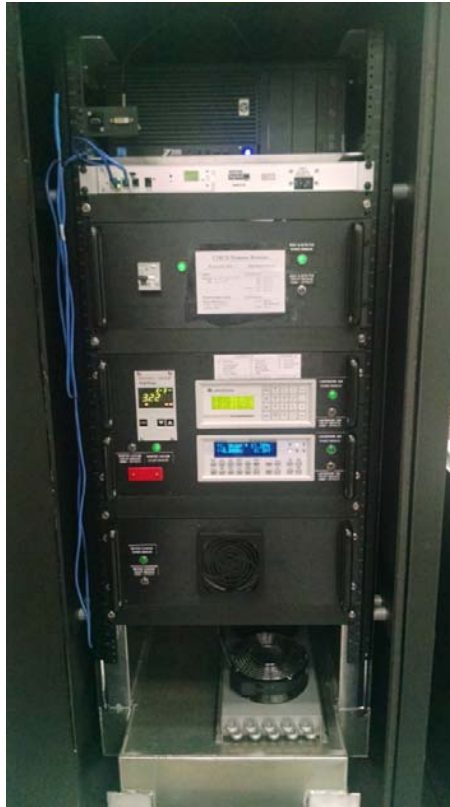

Figure 5 - View of the CIRCE electronics thermal enclosure, with the front door open.

## 6. CIRCE Software Systems

The CIRCE software control system aims to (a) control and monitor the array bias, clocking, and exposure time, (b) control and monitor the various mechanisms, (c) control and monitor the environment, (d) exchange information with the GTC's Telescope Interface Server (TIS), (e) display the image to the user and offer quick-look analysis tools, and (f) offer a complete data pipeline to produce science-quality data. These aims are accomplished through a three-tiered software system comprised of low-level agents that talk to the hardware, high-level Java GUIs that the user interacts with, and a python-based data reduction pipeline. We describe each of these tiers here.

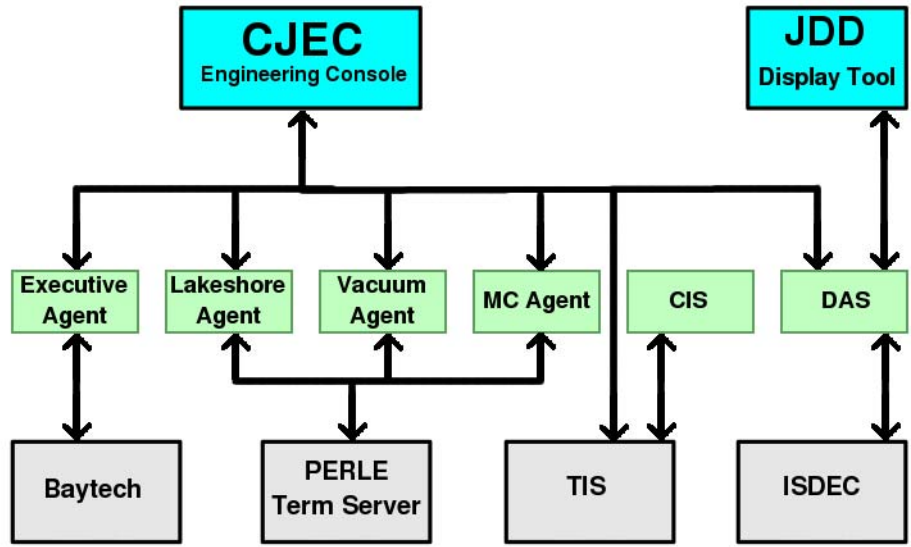

Figure 6 - CIRCE control software layout.

### *6.1 CIRCE Agents*

The CIRCE agents run on the CIRCE LCU and communicate directly with the detector, various devices, and the TIS.

### *6.1.1 MC agent*

The Motor Control (MC) agent communicates with the motor indexers via the terminal server that routes Ethernet ports to direct serial I/O ports.  The MC agent accepts network connections from clients such as the CIRCE Java Engineering Console (CJEC - see below) and allows commanding and monitoring the motor indexers.  The MC agent supports either low-level stepping commands or high-level positioning or datum (homing) commands, where the position step counts are configured in a file that is read by the MC agent, and editable by users.  The MC agent is written in C++ language.

### *6.1.2  DAS*

The Data Acquisition Server (DAS) communicates with the SIDECAR via the USB interface provided by the ISDEC. The DAS accepts network socket connections from client programs such as CJEC and the Java Data Display (JDD - see below) to configure the detector, initiate, and monitor the exposures. All facets of the detector array biases, clocking, readout modes, and exposure parameters are controlled via commands and queries sent to the DAS by clients. The DAS then converts commands to SIDECAR transaction packets, which are sent to the ISDEC via USB, and the ISDEC forwards the packets onto the SIDECAR communication interface, and reads replies from the SIDECAR, sending replies back to the DAS, which is then forwarded to clients. Note that the SIDECAR microcode is always looping over frames, either resetting or reading them, so when requested by a client, the DAS starts an exposure by commanding the ISDEC to synchronize with the SIDECAR to receive digital frame readout data. The ISDEC then forwards the readout frames to the LCU via the USB connection.  As each pre-allocated LCU USB memory buffer fills up, the DAS grabs each readout frame and writes the frame to the MEF (Multi-Extension FITS) data file on the LCU disk drive. Up to 100 clients are allowed connect to the DAS, so many clients can be either remotely monitoring exposures or initiating them. This has been helpful in allowing CIRCE team personnel to monitor GTC observations from Florida, for instance, and provide remote support for GTC astronomers. A stream of frame counter messages is sent to all JDD clients, providing notification of available data, and the DAS processes each ramp difference between the final minus starting frames and sends the difference images to clients, such as JDD, upon request. The DAS is a multi-threaded process, written in C++, and creates a server thread for each client connection, and has separate threads for grabbing, processing, and writing the data frames, all running asynchronously.

Using CJEC, the DAS can be commanded to perform exposures in either FS (Fowler Sampling) or URG (Up the Ramp Groups) with either full-frame readouts or sub-array readouts. Frame resets at the beginning of each ramp can be either fast (40 ms) or in normal frame time. Sub-array readouts can be either a horizontal band of pixels consisting of any number of full-rows, or an arbitrary window of pixels. Observations can be performed as a sequence of ramps of readout frames that are written to the data file as acquired, or in "burst" mode for sub-array readouts, during which readout frames are grabbed as fast as possible and kept in DAS memory, and written to disk after the observation is complete, allowing the CPU to easily keep up with many readouts per second. For example, by selecting the pixel clock of 200 KHz and specifying a band of 2048 columns by 10 rows the burst mode can run at 250 readouts per second, and faster if a smaller band is used.

### *6.1.3 CIS*

The CIRCE Interface Server (CIS) provides the interface between CIRCE control system DAS and the CORBA-based GTC control system Telescope Interface Server (TIS). The CIS handles transactions for obtaining FITS header information about the telescope systems by accepting requests from the DAS and then invoking CORBA methods that transact with the TIS, then sending the information back to the DAS. The CIS also copies each data file from the LCU disk to the GTC archive disk and notifies the TIS when data is ready. The CIS is written in Java.

### *6.1.4 Lakeshore and Pfeiffer agents*

The Lakeshore agent communicates with both the Lakeshore LS-331 temperature controller and Lakeshore LS-218 temperature monitor via socket connections and the Vacuum agent connects to the Pfeiffer pressure monitor, providing complete monitoring of environmental conditions. Meanwhile the Executive agent communicates with the Baytech RPC3A-16 AC power control unit and is able to stop and restart any of the other agents.

### *6.2 CIRCE Java Engineering Console (CJEC)*

The CIRCE Java Engineering Console (CJEC) is the primary user-interface to run CIRCE. It is a Java-based GUI that can be run on any machine on the same network with the CIRCE LCU. It communicates with all of the CIRCE agents, as well as the TIS, via TCP socket connections. The CJEC main tab contains a full sequencer that can be used to load and run observation sequences designed using GTC's Phase II web based observation planning tool. It also contains panels displaying agent statuses, current temperatures, and providing high level detector controls. CJEC also contains 6 low-level tabs for advanced users that allow full control and monitoring of the detector, motors, telescope dithering, power management, and the environment. We present a screenshot of the CJEC (set to its "Main" tab) in Figure 7 below.

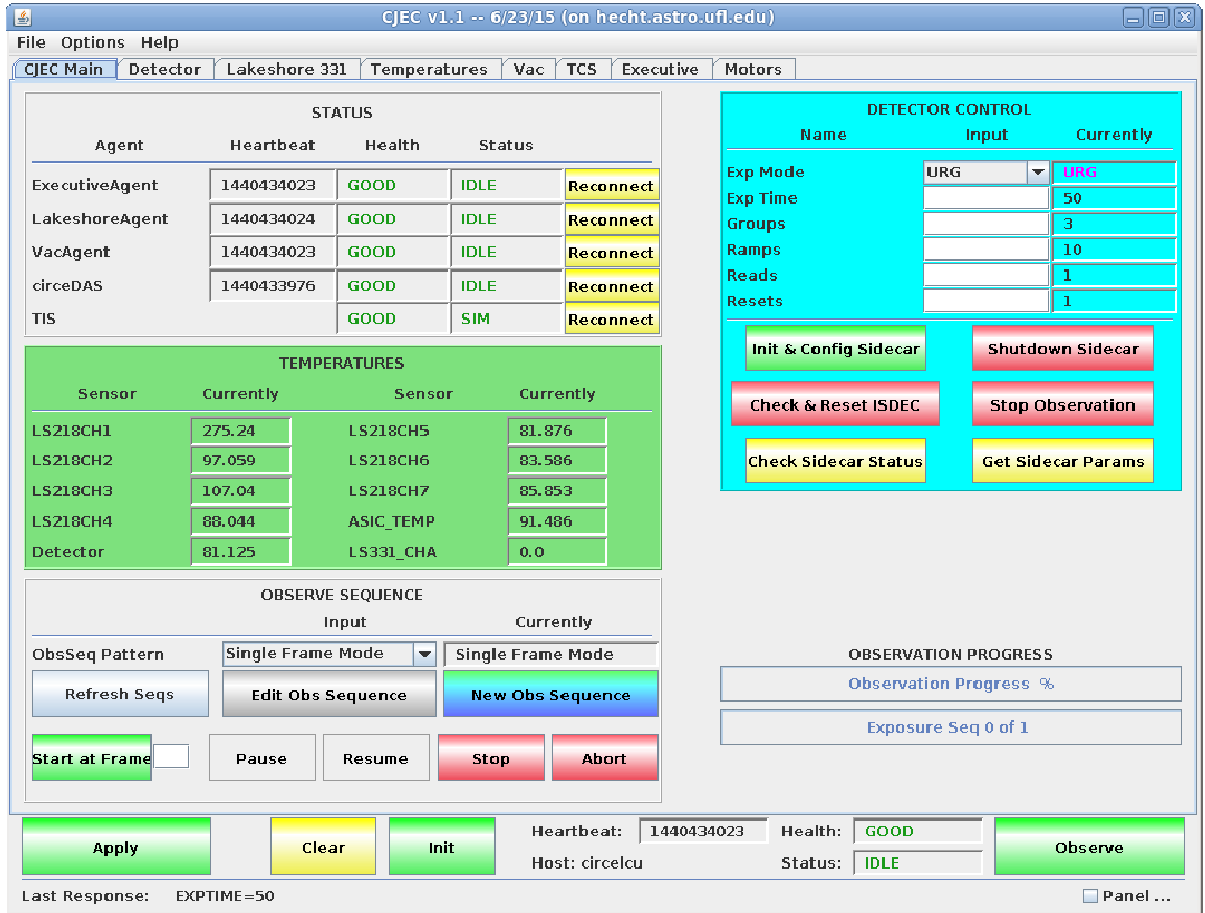

Figure 7 - CJEC screenshot

### 6.3 Java Data Display (JDD) tool

The Java Data Display (JDD) tool can be run on any workstation and connects to the DAS to monitor and display the exposures and detector readout images. JDD receives a stream of frame counter notification messages from the DAS during exposures, and when each ramp of frames is processed the JDD requests the resulting difference image from the DAS. Displayed images are thus the difference between the final frame minus the initial frame of each ramp. The JDD GUI consists of two "frames": one window to show the full image and select a region to zoom, and second window to display the zoomed region with some user controls for image scaling and analysis, such as line cuts, statistics in a box, and radial profile plots. JDD can also connect to the Telescope Interface Server (TIS) when JDD is used on-site at the GTC, and then provides a special panel to control telescope offsets and focus. A special option in JDD allows the user to select a point in the zoomed region and request a telescope offset that will move that point to the center of the FOV for subsequent exposures. JDD also has an expose button with a field for specifying the exposure time in seconds, but all the other exposure parameters are specified using CJEC. There is also a button to request and display the current FITS header of exposure data files, and in addition, buttons that will command the DAS to initialize the SIDECAR (loading the ASIC microcode) or perform shutdown (power off the detector).

### *6.4 superFATBOY data reduction pipeline*

The UF-developed superFATBOY data reduction pipeline [Warner et al., 2015, in prep; Warner et al. 2013, ASPC, 475, 79] is fully compatible with CIRCE data. superFATBOY is a python-based next-generation data pipeline that can be configured and extended to reduce data from virtually any near-IR or optical instrument. It can be installed on any Linux or Mac based computer and is run from the command line by providing an XML configuration file that describes (a) the data, (b) the processes that will be applied to the data, and (c) global parameters. Additionally, superFATBOY is designed to be able to harness the power of massively parallel algorithms developed using Nvidia's Compute Unified Device Architecture (CUDA) platform to enable near real-time data processing on machines equipped with a CUDA-enabled GPU (while retaining the ability to run on other machines in "CPU-mode" by simply changing one parameter in the XML configuration file). We have developed CIRCE-specific recipes as part of superFATBOY and have used superFATBOY to reduce the data presented in the images shown in Figures 12-13elow.

### 7. Laboratory Testing

Prior to delivery of CIRCE to GTC for on-telescope installation, we carried out a series of tests – both warm and cold – to verify and characterize the instrument performance. We describe some of the key tests and their results here.

### 7.1. *Mirror surface roughness tests*

The first test that we performed on the optical system of CIRCE was a measurement of the surface roughness of each individual mirror. Since the optical system was already aligned by the manufacturer, we chose a non-invasive technique to measure the surface roughness of each component to avoid having to remove any of the mirrors from the bench. The surface roughness of the mirrors is not necessarily a critical element for the image quality as long as the nominal specifications of the overall system are fulfilled. CIRCE's specification for the surface roughness of the collimating and imaging mirrors is <10 nm RMS, and <7.5 nm RMS for the two flat fold mirrors.

We used a very simple technique to measure the surface roughness without disrupting the optical setup. We took images of the reflected diffraction patterns produced by the mirrors of a HeNe laser beam. We estimated the surface roughness using Marechal's approximation. The Marechal approximation uses values of the Strehl ratio to obtain the surface roughness RMS in waves with errors of ~10% amplitude. The Strehl ratio is a measurement of the optical quality of a system, calculated as the ratio of the amount of light contained in the Airy disk of the diffracted image vs. a theoretical perfect maximum. The expected nominal surface roughness for each individual mirror is between 7.5 and 10.0 nm RMS, with a total surface roughness for the optical system of < 26.5 nm RMS.

Our measurements confirmed that the mirrors Collimator 1 and 2, and Imager 2, 3, and 4 meet specifications; Imager 1 was marginally above specifications; and the two Fold mirrors were well above the expected values. The combined surface roughness of all mirrors was initially 30±3 nm RMS, slightly above the nominal specifications for the complete system. We confirmed our results of the surface roughness of the two fold mirrors using a profilometer located at the laboratories of the Department of Mechanical and Aerospace Engineering at the University of Florida. The profilometer showed surface roughness values in the range 15 to 25 nm RMS for both fold mirrors. Based on these results, we removed the two fold mirrors and sent them back to the manufacturer for re-finishing. The final RMS surface finish for these mirrors was improved to <7.5 nm RMS, and the overall system surface RMS was ~23nm RMS - well within the specifications.

*7.2. Warm image quality tests*

The fabrication of the mirrors, brackets, and bench using the same material produces a homologous contraction of the optical system, so that the properties of the system are conserved at room and cryogenic temperature. This allows testing/verifying the image quality of the entire system at room temperatures, and in the visible range of the electromagnetic spectrum. To perform the test, we placed a visible light source at the entrance of the optical system, and a 640x480 pixel SBIG CCD camera with a 5.6μm pixel size (much smaller than the 18 μm IR array pixels) at the instrument focal plane location. The pixel scale of the CCD camera was ~0.03 arcsec/pixel (much smaller than the 0.1 arcsec/pixel of CIRCE's IR array), covering a field of view (FOV) of 0.32x0.23 arcmin, and thus insufficient to cover the 3.4x3.4 arcmin FOV of CIRCE instantaneously. In order to be able to test the whole FOV, we mounted the CCD camera on a two-dimensional computerized translation stage – lateral translation and focus translation – and manually moved the camera in the vertical direction. The tests were carried out using a pinhole mask located at the telescope focal plane. The mask contains 37 pinholes spatially-distributed over the entire 3.4x3.4 arcmin FOV. Each pinhole has a diameter of 170 μm, simulating 0.2 arcsec (2 pixels of the IR array) stars.

CIRCE's optical design is optimized to reduce the system aberrations to a minimum. ZEMAX simulations of the optical layout show astigmatism as the dominant source of aberrations. We analyzed the optical system taking through-focus images of several pinholes, covering a range of 200 μm along the optical axis centered in the best-focus position, with increments of 20 μm. Fig. 8-3 shows the through-focus images of the central pinhole. As expected from the simulations, astigmatism is the dominant source of aberrations. The elongation of the spot changes directions as the images pass through the focal plane, and is minimum and negligible at the best-focus location. The GTC is designed as a Ritchey-Chretien telescope with significant positive field curvature. The positive field curvature means that the focal surface is curled "up" at the edges, or in other words, the focal point at the field center is further along the optical axis than at the field edge. CIRCE is designed to compensate/flatten the field curvature onto the detector, i.e., CIRCE optics produces a field curvature, equal in amplitude, but opposite in direction to the telescope field curvature. The CIRCE design has a nominal on-telescope field curvature between 30 and 60 μm. A field curvature of 40 μm results in a defocus blur of »0.05 arcsec FWHM at the worst cases. The geometry of our test setup (pinhole mask) produces a flat field at the entrance side of the CIRCE optics. Thus, the output of the tested system is not a flat plane, but a curved focal surface where the edge of the field is ahead of the center along the optical axis. The ZEMAX simulations of the isolated CIRCE system with a flat pinhole mask to feed the instrument, result in a center to edge defocus of 250 μm. We used the CCD camera mounted on the translation stages to obtain the best-focus position of all the images of the pinholes in 5 different rows of pinholes. From top to bottom, we measured rows 9, 10, 11, 13, and 15, which cover approximately half of the focal surface. Fig. 8-4 shows the measurements of the best-focus location for the images of the pinholes. Our results are consistent with the ZEMAX simulations. They show the expected curvature of the focal plane, with a maximum in the center of ~250 μm. Thus, our results confirm that CIRCE will compensate for the positive curvature introduced in the image by the Ritchey-Chretien design of the GTC.

CIRCE's optical design is optimized to get the most out of the good sky quality at the GTC site, best-seeing of 0.3 arcsec in the IR. The high specifications imposed on the CIRCE design will ensure that CIRCE delivers seeing-limited images, even under the best seeing conditions at the GTC site. CIRCE's specifications impose an upper limit in the quality of the images lower than the best-seeing at the GTC site, i.e., CIRCE's image quality is always seeing-limited. The most important test performed on the optical system quantifies the FWHM of 0.2 arcsec simulated stars in order to analyze the quality of the images taken by our instrument. We measured the FWHM of several of the 0.2 arcsec pinholes images distributed around the entire FOV. We found that the FWHM values of the images range from 0.2 to 0.25 arcsec across most of the detector, with a slight increment (0.334 arcsec) in the bottom-left corner. The center and top images have excellent quality. Assuming a best seeing of »0.3 arcsec FWHM, we would expect delivered image quality of 0.33 arcsec over 90% of the array. That is only a 10% degradation of the image. Table 8-2 shows the values of the FWHM of the 0.2 arcsec pinhole images at 9 different positions spread over the FOV. It also shows the expected values for a best seeing of 0.3 arcsec. Fig. 8-5 shows the 0.2 arcsec pinhole images used to measure the image quality. From the FWHM test, we conclude that the image quality of CIRCE is very good, if not quite to the nominal CIRCE goals, especially in the Bottom-Left Corner. We note that these tests were performed with the original (poor surface roughness) fold mirrors, and the locations of poorest image quality were correlated with the regions of worst surface roughness. We conclude that this value of surface roughness in the Bottom-Left Corner of the Fold mirror 1 explains the degradation in the image quality of that portion of the final image.

### 7.3. *Warm flexure tests*

The final test performed on the optical system of CIRCE quantifies the flexure of the optical system with the rotation of the instrument. We rotated the optical system in increments of 45° until we completed a full rotation. We took images of several pinholes in each position and compared the series of images of each pinhole along the rotation. We found no deformation of the pinhole image shapes with the rotation angle, thus guaranteeing conservation of the optical quality of the system with the rotation of the instrument. We found shifts of the image location on the array of up to ~200 μm in a full lap. Those maximum shifts correspond to ~1 arcsec on the sky for a 360-degree rotation of the instrument. In a standard observation with CIRCE, an extremely long single run will be of ~30 minutes, that is equivalent to a maximum rotation of the instrument of 7.5-degrees and a shift of the image in the array of 4.1 μm, i.e., approximately a quarter of a pixel (18 μm) or a quarter of the 0.1 arcsec/pixel plate scale. Therefore, we conclude that the shifts of the image in the array with the rotation of the instrument are negligible during real observations.

### 7.4. *Laboratory cold tests*

In late 2013 and early 2014, after the arrival of the final hardware needed for operating the CIRCE HAWAII-2RG detector, we carried out thorough testing of the instrument at is cryogenic operating temperature ("cold testing"). The first set of tests focused on the cryovacuum performance of the system. CIRCE has 9 temperature sensors distributed throughout the dewar interior, allowing monitoring of key components (i.e. Detector and SIDECAR array controller) as well as gradients across the optical bench. All temperatures were found to reach their nominal design goals, with excellent thermal stability. We operated the detector without active control/heating, and find that its temperature is stable to ~ 0.001 K over typical integration times with CIRCE. We also found that the CIRCE vacuum pressure typically reaches ~1-2 microTorr after its initial cooldown, with long-term stability at those levels (varying slightly on long timescales with differences in ambient temperature as well as atmospheric pressure which alters the boiling point of the liquid nitrogen reservoir).

During this same period, we also carried out extensive reliability testing of the primary CIRCE cryogenic mechanisms. For the pupil mechanism, we executed >1500 motions over a full range of gravity vectors (by re-orienting the dewar in the handling cart). For the linear slide mechanism, we carried out >300 motions over the full range of gravity vectors. All of the test motions completed successfully, with no mechanism failures.

We then carried out image quality and flexure deflection tests using pinhole images - as for the warm tests described above, but this time under cryogenic conditions and using the CIRCE HAWAII-2RG detector. We measured the flexure impacts on CIRCE by rotating the dewar through a full range of angles in the instrument handling cart, taking pinhole images (including Hartmann masks) at 45-degree intervals. The tests were done with the CIRCE bench perpendicular to the gravity vector at 0 degrees - this is the worst case orientation for flexure.

Flexure has two overall impacts on instrument performance. The first, due to "internal flexure", is the change in alignment between CIRCE's internal components - in particular the optical components. Flexure-induced mis-alignment can lead to potentially significant image quality degradation if internal flexure is present. The second flexure impact is "global flexure" - the change in alignment between the instrument and the telescope, which can cause image motion on the instrument detector. CIRCE was designed to have minimal internal flexure, keeping high image quality in all orientations, and to have reasonable global flexure, so that image motions during an imaging exposure are small (<0.5 pixels, with a goal of <0.1 pixels, during a 1-minute maximum exposure). We note that for NIR seeing-limited imagers, long-term image motion on the scale of 10s of pixels is generally not important - the need for on-sky dithering means that the source locations will be changing by 10-100 pixels on ~1-minute timescales in any case.

We show the impact of flexure on CIRCE image quality in Figure 8 below. As can be seen, CIRCE maintains excellent image quality (<0.3 arcsec FWHM) in all orientations. Thus, CIRCE meets the goals here.

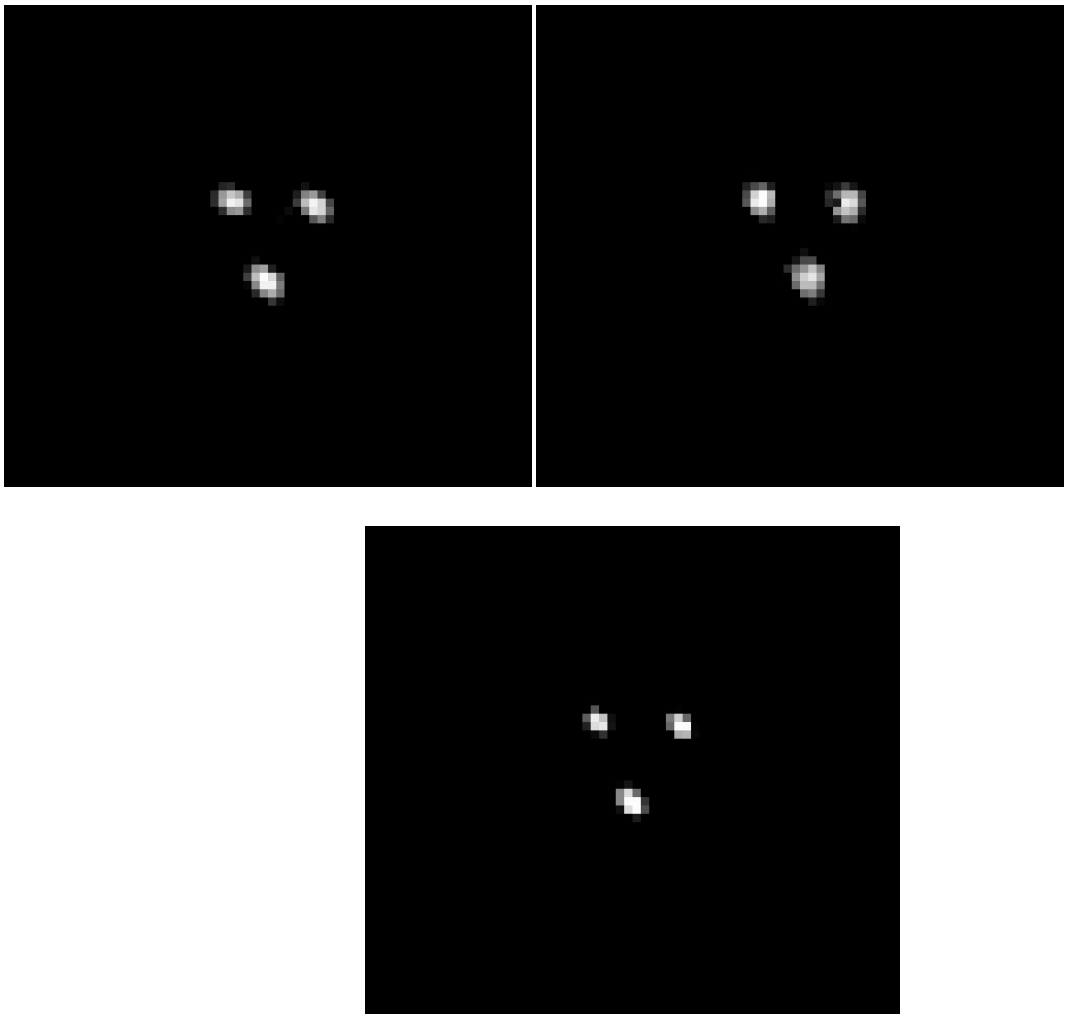

*Figure 84* - *CIRCE Ks-band pinhole images at a range of orientation angles. Each group includes images at 0-degrees, 90 degrees, and 18 degrees. Images have FWHM values ranging from 2-3 pixels (0.2-0.3 arcsec), with input pinholes of 2.0-pixel (0.2-arcsec) diameters.*

We present the impact of flexure on global image motion in Figure 9. As we can see, the maximum image motion is much less than 0.1 pixel/exposure for the longest expected exposure time for imaging (60s).

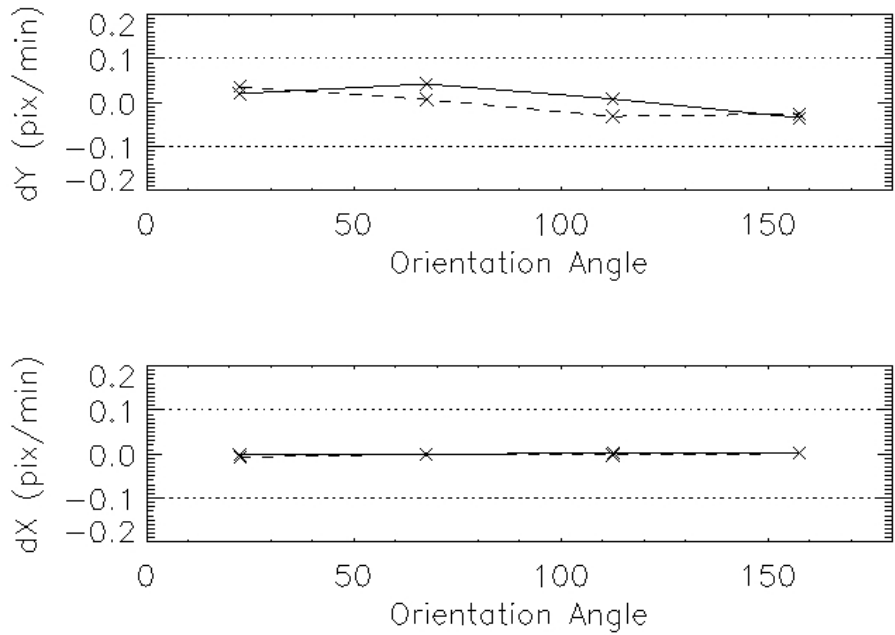

*Figure 9* - *CIRCE image motion versus orientation angles. We note that the largest rate of motion is <<0.1 pixels during a long (1-minute) exposure. Typical CIRCE exposures will range from ~5s to ~12s (see Table 3), so this plot shows an extreme "worst case" performance - and even then CIRCE exceeds the performance goal significantly.*

As a broadband NIR ground-based imager, CIRCE is relatively insensitive to detector read noise, due to the large intrinsic sky background. The nominal requirement is that reasonable scientific exposure times produce larger shot noise from incident photons than the read noise (i.e. background-limited or target-limited performance). While target-limited performance varies with target, the background-limited case actually provides the fundamental limit which read noise needs to reach (i.e. for the faint-target case). For CIRCE, we present the expected sky backgrounds per pixel as a function of filter band exposures in Table 3, along with the corresponding read noise requirement.

**Table 3 - CIRCE Sky Background and Instrument Noise**

| Bandpass | Sky Bkgd (e-/s/pix) | Typ. Integration Time (s) | Sky Flux (e-/pix) | Sky Noise (e-/pix) | Read Noise (e-) | In-dewar Bkgd (e-/pix/s) |
|---|---|---|---|---|---|---|
| J | 400 | 5 | 2000 | 44.7 | 50 | 1.3 |
| J | 400 | 12 | 4800 | 69.3 | ~20 (Fowler8) | 1.3 |
| H | 1280 | 5 | 6400 | 80 | 50 | 1.3 |
| Ks | 2700 | 5 | 13500 | 116 | 50 | 1.3 |

The currently achieved readout noise with CIRCE's engineering-grade HAWAII-2RG detector is ~50e- RMS for a single read in a CDS operation (see Figure 10). In laboratory testing, we have found that multiple non-destructive reads (Fowler sampling) produce effective read noise approaching <20 e- RMS , with the added constraint that the minimum read time increases from 1.5s for CDS to 12s for Fowler8 sampling. We present these results in Table 3 as well, confirming that CIRCE is capable of providing background-limited performance for scientifically reasonable exposures times in all bands.

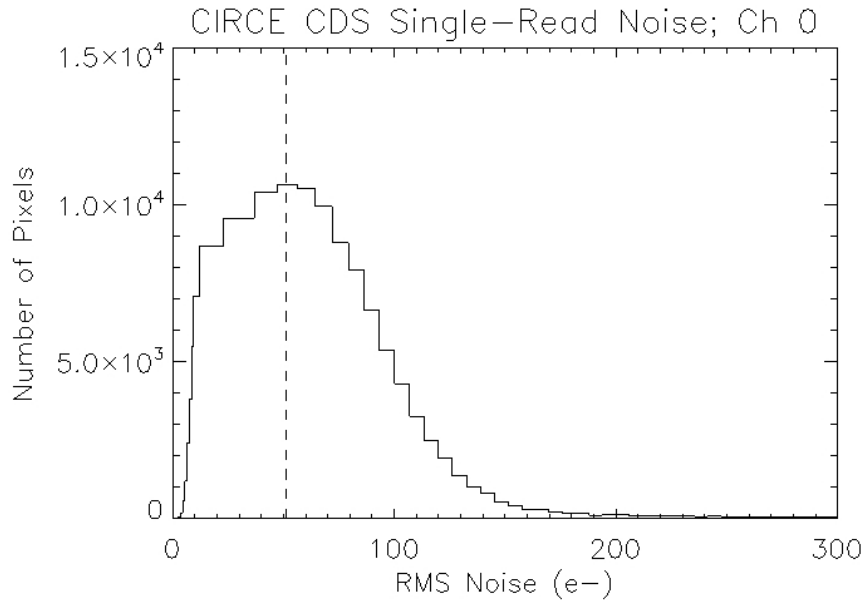

*Figure 10 - CIRCE read noise for channel 0. The vertical dotted line indicates a value of 50 e-.*

We note that this performance is worse than the nominal performance for a science grade device (<20e- RMS, with ~12e- RMS goal are the manufacturer's claims). The exact source of this read noise is unclear, but may be due to the CIRCE engineering grade device itself - Teledyne has confirmed that this device does not meet the science-grade performance requirements, as expected for an engineering-grade device. We also note that there is "odd-even" channel striping in the noise. Furthermore, we compared the device performance using both the ISDEC controller interface as well as a Teledyne JADE2 card in the laboratory. We found that the ISDEC matched the JADE2 noise performance (in fact, there is some weak evidence that the ISDEC card is slightly better than the JADE2 in terms of readout noise performance).

We present the in-dewar background measurements for CIRCE in Figure 11 below. We took these data after installing internal baffling to minimize light leaks from the dewar exterior. These data were taken with a 9dB gain setting on the SIDECAR controller, providing a gain of 8.1 ± 0.2 e-/ADU in the system. (We note that the gain was later reset to ~5.4 e-/ADU for on-telescope use). Thus, the slope of the plot in Figure 11 corresponds to an in-dewar background of 1.3 e-/s/pixel. This level is well below the sky background for CIRCE's lowest-background filter bandpass (J-band) - see Table 1 - and we conclude that CIRCE meets the scientific requirements for background-limited observations in all bands.

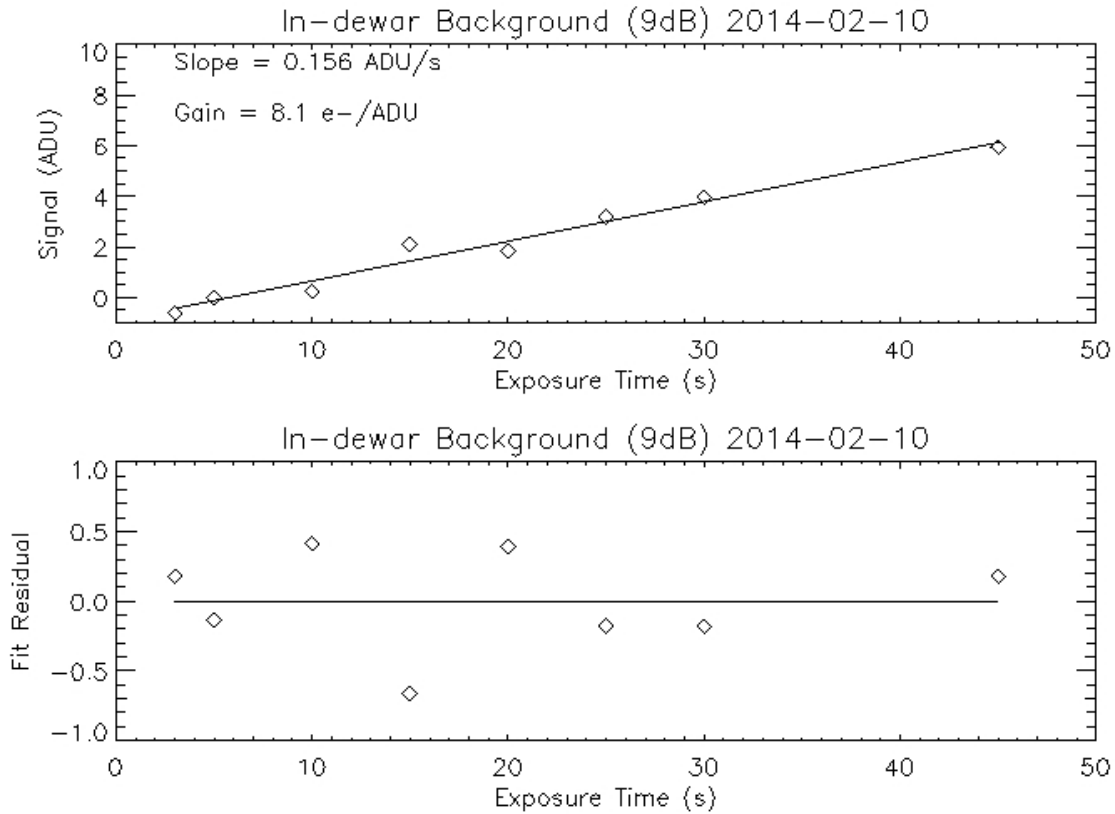

*Figure 11 - CIRCE in-dewar background. (Top) Detector signal versus exposure time. (Bottom) Residuals to a linear fit with a slope of 0.156 ADU/s.*

## 8. CIRCE Installation and Performance at GTC

### 8.1. *Arrival and laboratory test at GTC*

CIRCE arrived at the GTC telescope facility in late June 2014, and the UF CIRCE team traveled to La Palma and unpacked it in July 2014. We found the instrument subsystems and components to be in excellent condition, with no shipping damage. Over a period of 8 days, the UF team, assisted by GTC staff, cleaned and re-assembled the instrument, culminating in a pumpout and cooldown of the instrument. We configured the instrument in its pinhole mask configuration, in order to test the detector alignment with the focal plane after its removal and re-install post-shipment. We found a slight tilt to the detector with respect to the optical axis, resulting in a small degradation in image quality due to defocus effects. Our analyses showed that this effect would be negligible for seeing worse than ~0.4-0.5-arcsec FWHM, and would have a small effect even for seeing as low as 0.35-arcsec. Thus, after consultation with the GTC staff, we decided not to take the time/risk to further adjust the detector tilt. After this test was completed, we allowed CIRCE to warm up in order to switch from the lab-testing configuration to the operating configuration.

On a second trip in August 2014, CIRCE team members returned to GTC. We removed the pinhole mask and installed the HWP mechanism. We then pumped and cooled the instrument a second time at GTC. We verified the full functionality and observation-readiness of the instrument in the lab. At that point, the UF team members returned home to await the readiness of the GTC Folded Cassegrain focal station, leaving the GTC staff to monitor and maintain the instrument during the remainder of the cooldown process.

### 8.2. *Installation on GTC & engineering first light*

In December 2014, GTC had successfully installed the Folded Cassegrain rotator onto the focal station and tested its alignment and rotation. The joint UF/GTC CIRCE team then installed the instrument in one day, followed by a day of daytime operations tests on the rotator. Note that while the rotator was capable of rotating CIRCE under software control, coordination of the rotator motion with the telescope tracking had not yet been implemented in software. This limited the FC rotator to providing either no motion or a constant angular velocity during observations on this engineering run.

CIRCE obtained her first light images on the GTC on December 10, 2014. We first verified CIRCE's pupil alignment to the GTC, as well as the location of the center of rotation. While the temporary unavailability of software coordination between CIRCE's rotation on the FC rotator and the telescope tracking prevented full science commissioning at this time, we were able to obtain images in all 3 bands. GTC staff enabled this by providing the capability to first calculate the average angular velocity expected during an upcoming observation of a particular target, and then setting the FC rotator velocity to match that value. While not perfect, this approach proved more than adequate to demonstrate CIRCE's capabilities on the GTC. In particular, CIRCE provided excellent (~0.40-arcsec FWHM) seeing-limited images during relatively long (~30s) J-band exposures on one night of this run.

### 8.3. *Science commissioning and on-telescope performance*

We initiated science commissioning in March 2015, immediately after the FC Rotator software was completed and fully tested by GTC staff. We obtained full sets of commissioning data for the primary imaging modes of CIRCE, and we were able to finish the science commissioning in June and early July 2015, with GTC staff operating the instrument and telescope with remote monitoring and support from the CIRCE team at UF. We present some of the early images in Figures 12 to 13.

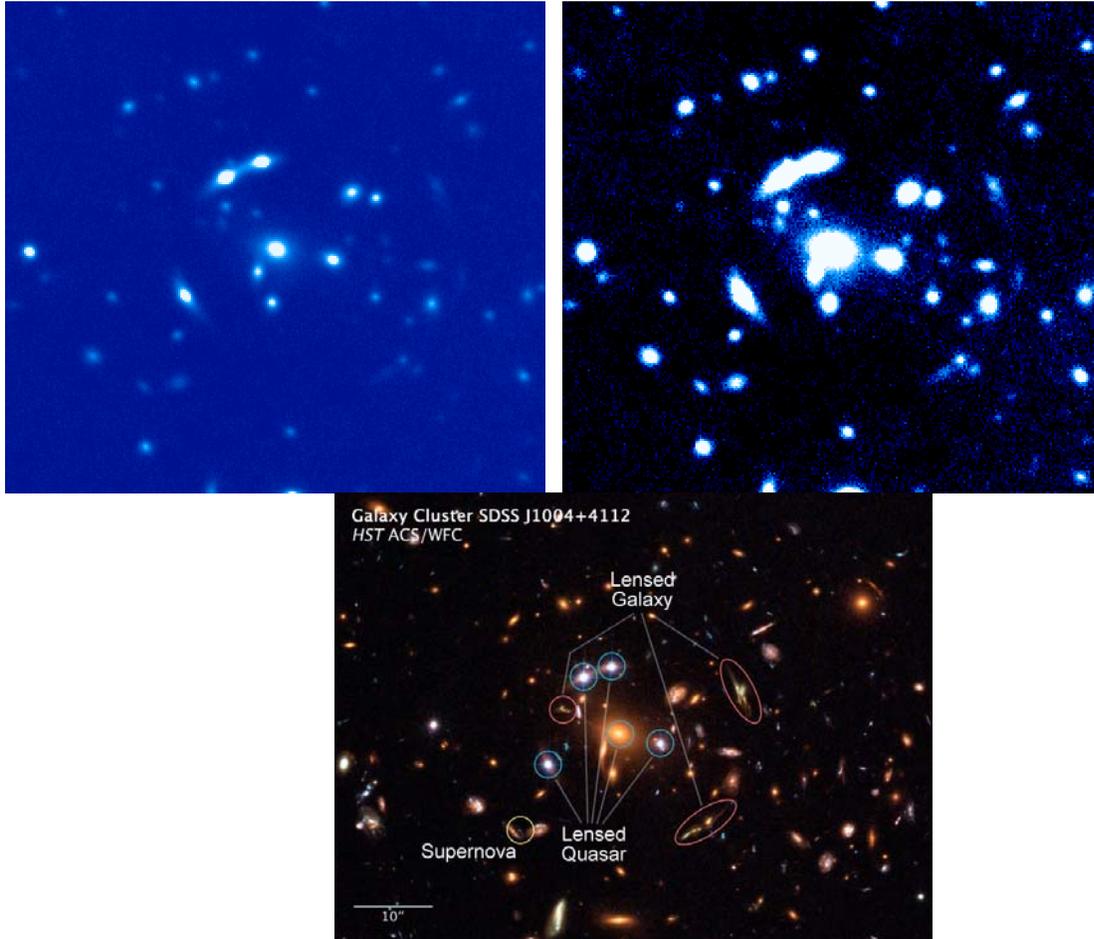

*Figure 12* - *Images of the lens system SDSS J1004+4112. (Top Left) Linear-scaled CIRCE Ks image. The image FWHM is ~0.40-arcsec. This image reaches a depth of Ks ~24.5 mag at 5σ, in about 90 minutes of on-source integration. (Bottom) HST image of the same field, with labels of key features for reference - image credits ESA, NASA, K. Sharon (Tel Aviv University) and E. Ofek (Caltech). This image was obtained as part of Proposal ID #10509 (Top Right) Log-scaled version of the CIRCE image, allowing a better view of fainter low surface brightness features.*

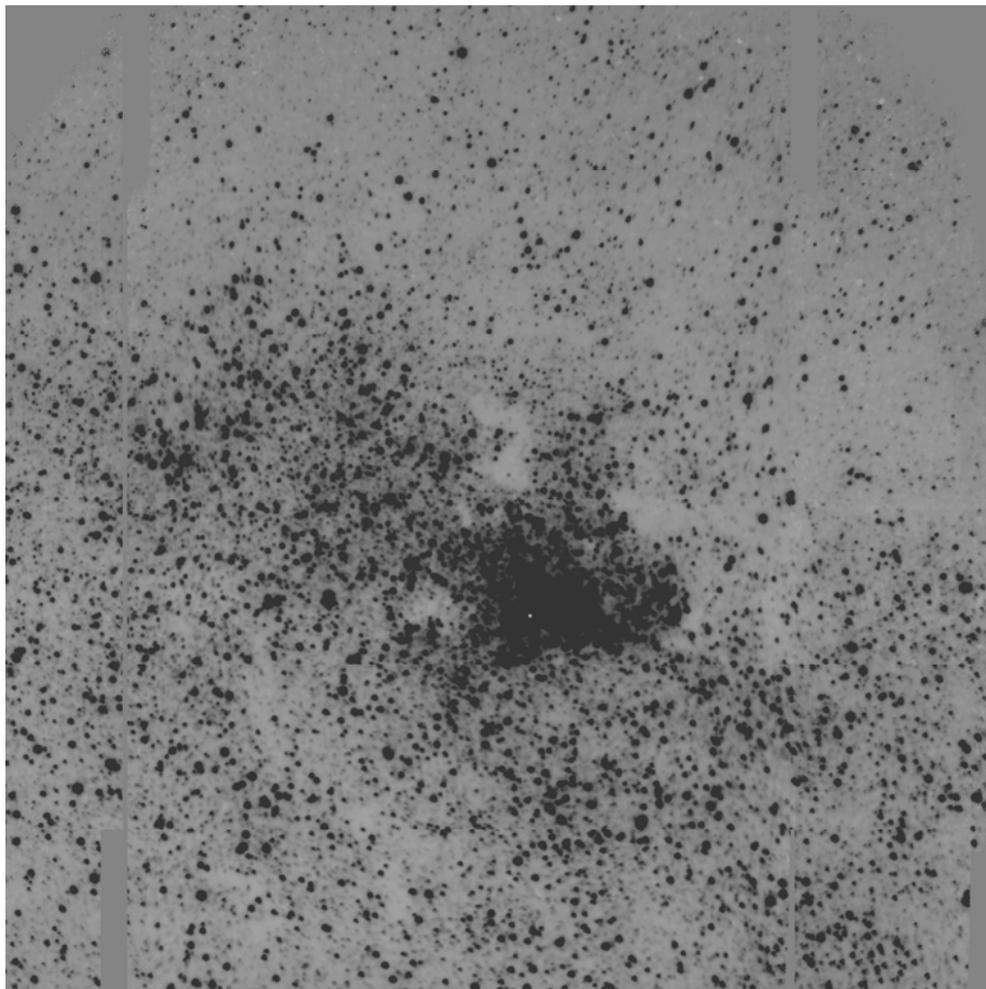

***Figure 13*** *– 4.2' x 4,2' CIRCE Ks image of the Galactic Center region taken July 2015. It is a stack of five dithered image with 15 seconds of total on-source exposure time. Seeing was 0.6" which is typical for the GTC.*

The primary conclusion of the commissioning effort is that CIRCE performance on-telescope matches the expectations based on laboratory testing. We present key on-telescope performance parameters below.

*8.3.1 On-telescope sensitivity*

In Table 4 below, we present the measured point-source sensitivity of CIRCE on the GTC. These measurements are for typical observing conditions at the GTC with the target at 1.2 airmasses, with S/N = 5σ for a 3600-s on-source exposure.

**Table 4 - CIRCE measured sensitivities on the GTC**
**(5σ, 3600-seconds, 1.2 airmass)**

| Band (Magnitude System) | 5σ Magnitude (0.8-arcsec FWHM) | 5σ Magnitude (0.4-arcsec FWHM) |
|---|---|---|
| J (Vega) | 24.0 | 24.7 |
| J (AB) | 24.9 | 25.6 |
| H (Vega) | 23.2 | 23.9 |
| H (AB) | 24.6 | 25.3 |
| Ks (Vega) | 22.5 | 23.2 |
| Ks (AB) | 24.4 | 25.1 |

*8.3.2 On-telescope image quality*

In Figure 14, we present the delivered image quality from CIRCE on the GTC, based on images taken in March 2015. The seeing was ~0.55-arcsec FWHM during these observations, and the measured point spread functions all have FWHM within ±0.1-arcsec of that value.

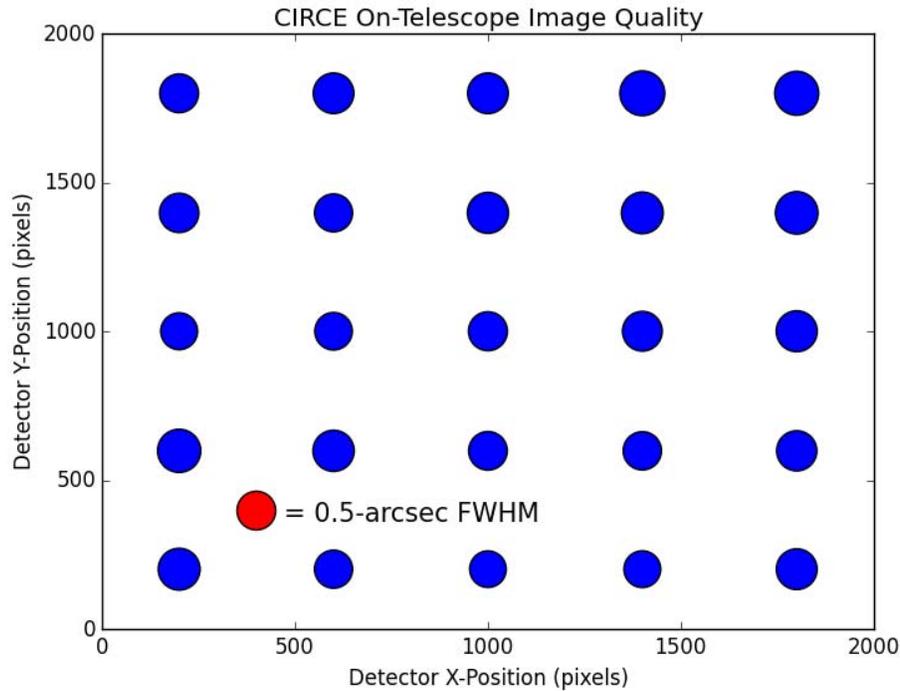

*Figure 14 - CIRCE delivered image FWHM versus detector position. The circle diameters are linearly proportional to the FWHM, and the red circle shows 0.5-arcsec FWHM for scale. The CIRCE PSFs in these seeing limited image have a median value of 0.54-arcsec FWHM.*

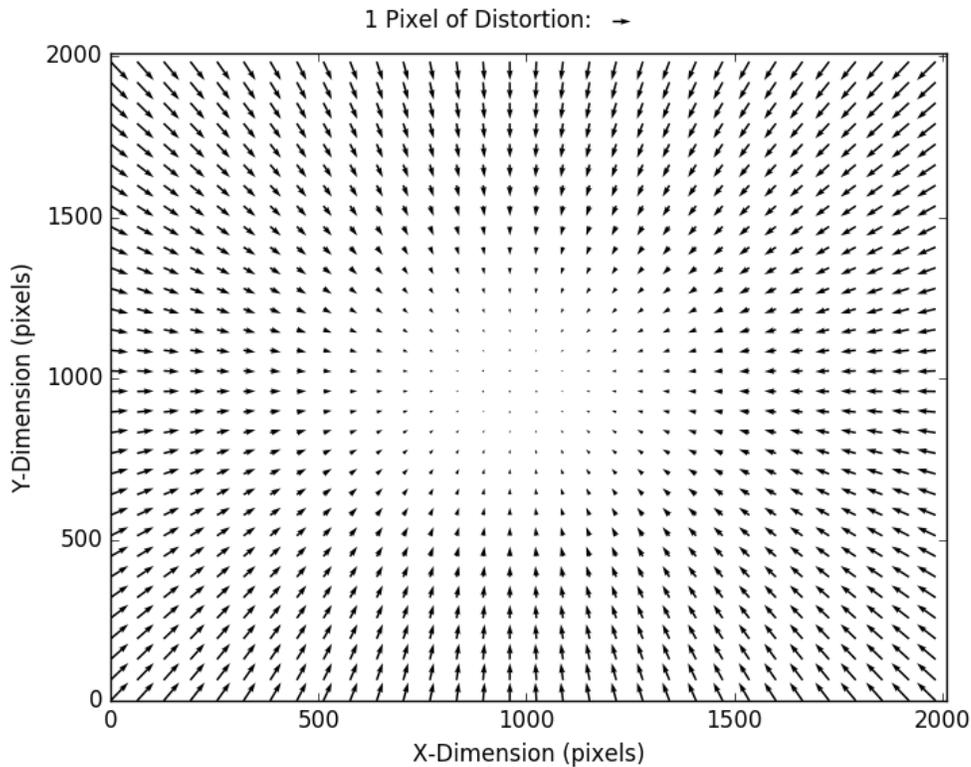

*Figure 15* - *CIRCE distortion map. The map above shows the magnitude and direction of the distortion in various regions around the image, though in reality the distortion correction is unique for each pixel. Note that the distortion is radial in nature, and the magnitude increases towards the edge of the field. Also worth noting is that the center of the distortion pattern (the region with the least distortion) is slightly off-center from the physical center of the array.*

As with most instruments, particularly those with off-axis aspheric surfaces, there is some inherent distortion in CIRCE. This distortion is relatively small, amounting to only 2 pixels at the corners, but can nevertheless be corrected using a 3rd-order 2-dimensional polynomial. This polynomial in turn is solved for each pixel in the frame, and the result for a given pixel tells you the magnitude and direction of its distortion. Figure 15, above, uses an arrow plot to help visualize the distortion correction across the field.

In order to generate the coefficients of the corrective $3^{rd}$-order 2-dimensional polynomial, we used known reference fields as astrometric standards. Such fields included the Sextans A dwarf irregular galaxy and the Galactic center region (500" from Sgr A* to avoid overcrowding). First, we ran the data through Source Extractor (http://www.astromatic.net/software/sextractor) to generate a source catalog for the field. This catalog included the location of each source in pixel-space for a given image. Next that catalog was fed into the SCAMP package (http://www.astromatic.net/software/scamp) which cross-referenced the sources in the catalog (with locations given in pixel-space) with 2MASS Ks-band astrometry of the field using pattern matching. Once it had matched the input catalog from CIRCE to the 2MASS catalog, it would then generate as part of its output a set of polynomial distortion coefficients needed to scale the CIRCE catalog to align with the 2MASS catalog.

When this process was completed for a single image, the distortion coefficients would be correct for that single image it was applied to. However, if the image did not have sources distributed across the entire array, then the solution could not be trusted for "dark" regions of the image where no sources were present to constrain the solution. Ideally, if this process was applied to a field that had stars uniformly distributed across the entire array, then that solution could be trusted as a general distortion correction for every image. In reality, when we applied this technique to images that were very crowded it often did not converge on a correct solution due to overcrowding and source confusion (i.e., the pattern matching would fail). Conversely, when we applied this technique to images that had relatively few sources, it would always converge on a solution, but as mentioned before that solution could not be trusted for the areas of the array that did not have a source nearby to help constrain the solution.

To solve this, we repeated this technique using dozens of images, and recorded the distortion coefficients for each one. For the images that were overcrowded and couldn't converge on a correct solution, the distortion coefficients stood out as obvious outliers and were discarded. The remaining solutions were all similar, and the median of each coefficient was taken to generate a new set of "master" distortion coefficients. To verify that the new solution was correct, we then applied this set of distortion coefficients to all of the images, and then repeated the whole process again to check for any uncorrected "residual" distortion. The results showed less than 0.3 pixels of distortion at the corners (the area of the array that shows the most distortion) in the worst case and essentially zero lingering distortion in the center of the image, demonstrating that our new set of distortion coefficients corrects for essentially all of the distortion present in the system. These distortion coefficients can be included in the image processing package superFATBOY as part of its image reduction pipeline.

### 8.3.3 On-telescope observing efficiency

A key parameter for the performance of any astronomical instrument is its "open shutter" efficiency - i.e. the fraction of wall-clock time during an observation for which the instrument is actively collecting photons. CIRCE on GTC performs extremely well in this category, with an open-shutter efficiency of ~75% for most reasonable observing patterns. This is substantially better than many other NIR instruments on 8-m to 10-m telescopes, which have observing efficiencies ~50% for similar observing strategies.

### 8.3.4 On-telescope polarimetry performance

We measured the CIRCE+GTC intstrumental polarizations as follows. We observed three null standards over two periods of polarimetry commissioning at different instrumental position angles (IPA) with rotator tracking disabled. By definition, we assume that a null standard has (or is close to) the intrinsic polarization state of [I/I, Q/I, U/I] = [1, 0, 0]. In an ideal case, the results of null standard observations should follow these equations as IPA ($\theta$) changes:

$$q = q_c + q_T \cos(2\pi(\theta/180°)) - u_T \sin(2\pi(\theta/180°)) = q_c + P_T \sin(2\pi(\theta/180°) + \phi_1)$$

$$u = u_c - u_T \cos(2\pi(\theta/180°)) - q_T \sin(2\pi(\theta/180°)) = u_c + P_T \sin(2\pi(\theta/180°) + \phi_2)$$

$$\phi_1 - \phi_2 = \pi/2$$

where $q_c$ and $u_c$ are the instrumental polarization of CIRCE, $q_T$ and $u_T$ are the instrumental polarization of the GTC telescope (with $P_T = (q_T^2 + u_T^2)^{1/2}$ ).

Standard dual-beam polarimeters are designed to obtain two of the parameters defined above ((q1, u2) or (q2, u1)) using 2 rotation angles of the HWP. CIRCE's double-Wollaston configuration allows measurement of pairs, (q1, u1) or (q2, u2), in a single exposure and HWP setting; however, the (q1, u2) and (q2, u1) pairs come from two different portions of the telescope pupil projected through the optical system of the instrument. Therefore, any polarization gradient of the components in the optical path across the pupil (such as the primary mirror of the telescope) will affect our results. To quantify this, we apply the calibration equations above on the pairs (q1, u2) and (q2, u1) separately.

We observed the following null standards during polarimetry commissioning:

1. GJ 3753 (J=11.6 mag, H=11.1 mag, Ks=11.1 mag)

2. BD+28 4211 (J=11.3 mag, H=11.4 mag, Ks=11.6 mag)

3. HD 331891 (J=8.8 mag, H=8.8 mag, Ks=8.7 mag)

For each filter, we select the (q1, u2) pair as a representative, and made the measurements of these parameters. Then, we applied the resulting calibration equations for (qc, uc, qT, uT) to arrive at model fits. We assume a 0.1% systematic uncertainty in each data point to account for differences in source polarization and atmospheric conditions. We performed two sets of fits - first with constant amplitude in the observed polarization, and second with an angle-dependent amplitude (which effectively allows for the pupil-dependent polarization affects noted above). We present these results in Figures 16, 17, 18 and Table 5 - $P_T$ refers to the $q_T$ measurements for variable-amplitude fits. As can be seen in Table 5, the typical systematic uncertainty in polarization is ~0.14-0.22% RMS, better than the requirement of <0.5% RMS and close to the design goal of 0.1% RMS.

**Table 5 - Instrumental Polarization Fits**

| Filter | Constant-amplitude Fit | | | | Variable-amplitude Fit | | | |
|---|---|---|---|---|---|---|---|---|
| | $q_c$ (%) | $u_c$ (%) | $P_T$ (%) | RMS (%) | $q_c$ (%) | $u_c$ (%) | $P_T$ (%) | RMS (%) |
| **J** | -1.04±0.04 | -0.05±0.06 | 0.97±0.06 | 0.29 | -1.04±0.04 | -0.04±0.06 | 0.90±0.08 | 0.22 |
| **H** | -1.03±0.03 | -0.13±0.04 | 0.86±0.04 | 0.17 | -1.03±0.03 | -0.12±0.04 | 0.84±0.05 | 0.14 |
| **Ks** | -1.17±0.03 | -0.26±0.05 | 0.95±0.04 | 0.19 | -1.17±0.03 | -0.26±0.05 | 0.94±0.06 | 0.19 |

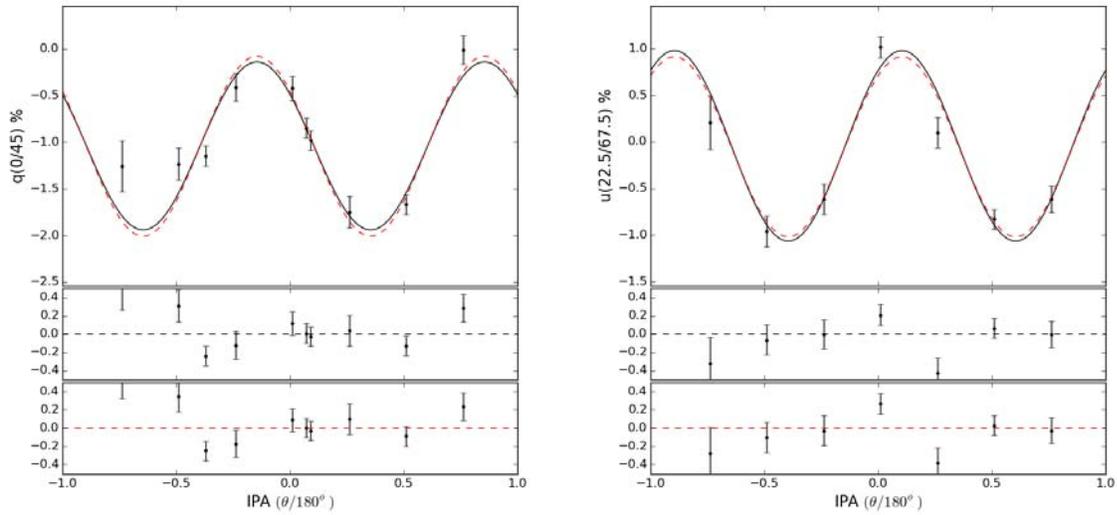

*Figure 16* - J-band instrumental polarization models. The black line shows the constant-amplitude fit, while the red line shows the variable amplitude-fit. Residuals for each model are shown in the bottom frames.

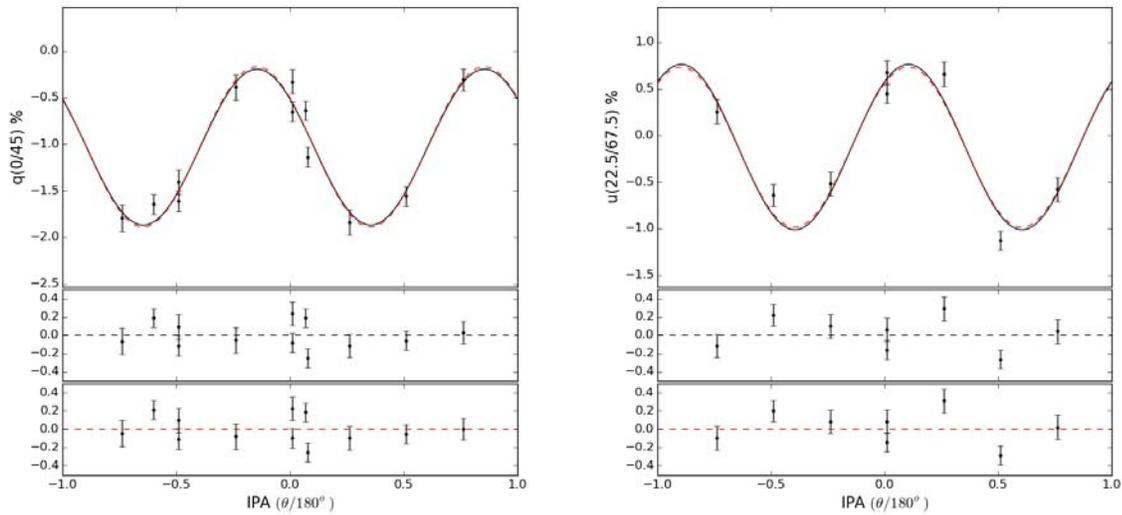

*Figure 17* - H-band instrumental polarization models. The black line shows the constant-amplitude fit, while the red line shows the variable amplitude-fit. Residuals for each model are shown in the bottom frames.

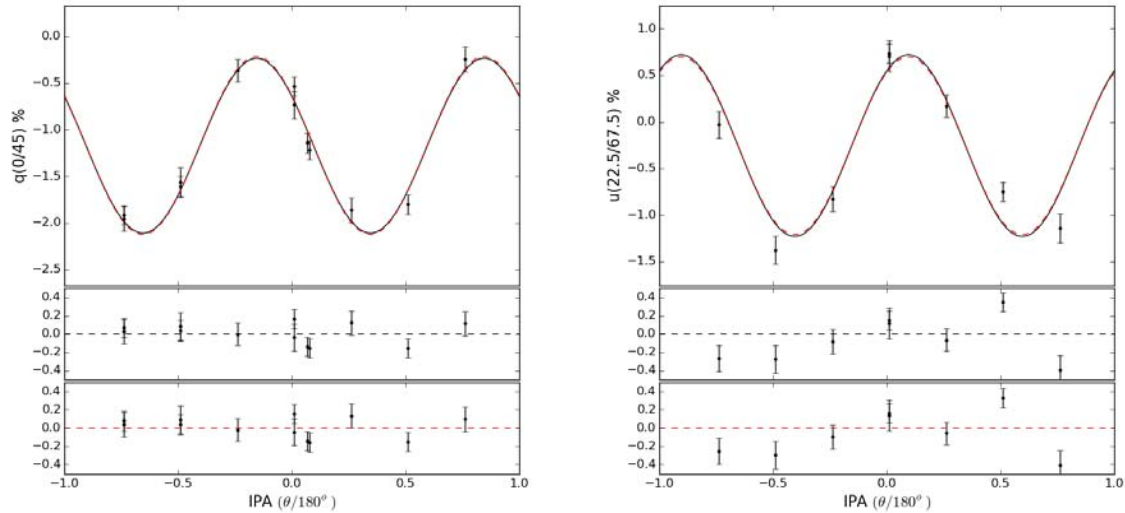

*Figure 18* - Ks-band instrumental polarization models. The black line shows the constant-amplitude fit, while the red line shows the variable amplitude-fit. Residuals for each model are shown in the bottom frames.

**Acknowledgments**

CIRCE was developed with support of the University of Florida and the National Science Foundation (NSF grant AST-0352664). The CIRCE team gratefully acknowledges the collaborative support of the Gran Telescopio Canarias management and staff in this endeavor - both the current staff and, in particular, the long-standing support of the previous Director Pedro Alvarez and the previous Project Scientist J.M. Rodriguez. (A.1)